\documentclass{aa}  
\usepackage{graphicx}
\usepackage{changes}
\usepackage{txfonts}
\usepackage{lscape}
\usepackage{natbib}

\begin{document}

\title{High-precision chemical abundances of Galactic building blocks.}
\subtitle{The distinct chemical abundance sequence of Sequoia}
\author{Tadafumi Matsuno \inst{1} 
\and 
Helmer H. Koppelman \inst{2}
\and 
Amina Helmi \inst{1}
\and
Wako Aoki \inst{3}
\and
Miho N. Ishigaki \inst{3}
\and
Takuma Suda \inst{4}
\and
Zhen Yuan \inst{5}
\and
Kohei Hattori \inst{3,6}
}

\institute{
   Kapteyn Astronomical Institute, University of Groningen, Landleven 12, 9747 AD Groningen, The Netherlands\\
   \email{matsuno@astro.rug.nl}
  \and
   School of Natural Sciences, Institute for Advanced Study, 1 Einstein Drive, Princeton, NJ 08540, USA
  \and
   National Astronomical Observatory of Japan, 2-21-1 Osawa, Mitaka, Tokyo 181-8588, Japan 
  \and 
 Department of Liberal Arts, Tokyo University of Technology, Ota-ku, Tokyo 144-8535, Japan 
  \and
 Universit\'{e} de Strasbourg, CNRS, Observatoire Astronomique de Strasbourg, UMR 7550, F-67000 Strasbourg, France 
  \and 
 Institute of Statistical Mathematics, 10-3 Midoricho, Tachikawa, Tokyo 190-0014, Japan
  }

\abstract
  {
Sequoia is a retrograde kinematic substructure in the nearby Galactic halo, whose properties are a matter of debate.
For example, previous studies do not necessarily agree on the chemical abundances of Sequoia stars, which are important for understanding its nature.
  } 
  {
  We characterize the chemical properties of a sample of stars from Sequoia by determining high-precision abundances. 
  } 
  {
  We measured abundances of Na, Mg, Si, Ca, Ti, Cr, Mn, Ni, Zn, Y, and Ba from a differential abundance analysis on high signal-to-noise ratio, high-resolution spectra from new observations and from archival data. 
  We compared precisely measured chemical abundances of 12 Sequoia candidates with those of typical halo stars from the literature, which also includes stars from Gaia-Enceladus.
  This allowed us to characterize Sequoia and compare it to another Galactic building block.
  The comparison was made after putting all the abundances onto the same scale using standard stars.
  } 
  {
  There are significant differences in [{Na}/{Fe}], [{Mg}/{Fe}], [{Ca}/{Fe}], [{Ti}/{Fe}], [{Zn}/{Fe}], and [{Y}/{Fe}] between Sequoia and Gaia-Enceladus stars at $-1.8\lesssim [\mathrm{Fe/H}]\lesssim -1.4$ in the sense that these abundance ratios are lower in Sequoia.
  These differences are similar to those seen between Gaia-Enceladus and in situ stars at a higher metallicity, suggesting that Sequoia is affected by type~Ia supernovae at a lower metallicity than Gaia-Enceladus.
  We also confirm that the low [{Mg}/{Fe}] of Sequoia is seen in the literature and in surveys, namely APOGEE DR17 and GALAH DR3, if the stars are kinematically selected in the same way. 
  } 
  {
  Sequoia stars have a distinct chemical abundance pattern and can be chemically separated from in situ stars or Gaia-Enceladus stars if abundances are measured with sufficient precision, namely $\sigma([\mathrm{X/Fe}])\lesssim 0.07\,\mathrm{dex}$.
  } 
\maketitle

\section{Introduction\label{sec:intro}}

The standard cosmological model predicts that galaxies grow through a hierarchical process.
The Milky Way is no exception and has been shown to have accreted a number of dwarf galaxies.
There is indeed evidence for the ongoing accretion in the form of stellar streams from the Sagittarius dwarf galaxy \citep[e.g.,][]{Ibata1994a,Belokurov2006a,Grillmair2006a,Bernard2016,Malhan2018a,Ramos2020a,Antoja2020a}.
However, in the case of ancient accretion events that deposited stars to the inner parts of the Galaxy, streams would have lost their spatial coherence.
This is why chemodynamical analysis of halo stars is a powerful way to recover the accretion history of the Milky Way.
Orbits and chemical abundances generally remain unchanged for a long time.

A number of studies have pointed out correlations between abundance ratios of halo stars and their kinematics \citep[e.g.,][]{Nissen1997a,Gratton2003a,Venn2004a,Nissen2010,Ishigaki2012}.
In particular, \citet[][hereafter NS10]{Nissen2010} clearly showed the presence of two chemically distinct stellar populations among nearby halo stars that also have different kinematics.
They interpreted the population with a low [$\alpha$/{Fe}] abundance ratio as a group of accreted stars from dwarf galaxies and the high-$\alpha$ population to have formed in situ within the Milky Way.
After the data releases from the Gaia mission \citep{GaiaCollaboration2016a,GaiaCollaboration2018a}, it became apparent that there is a kinematic overdensity of stars with radial orbits in the Galactic halo, known as the Gaia-Sausage \citep[e.g.,][]{Belokurov2018a,Koppelman2018a}.
The kinematic overdensity turns out to follow the low-$\alpha$ population \citep[e.g.,][]{Helmi2018a,Haywood2018a,Mackereth2019a,Hasselquist2021a} and it is now considered to be the debris from the last major merger that the Milky Way experienced and it has been named Gaia-Enceladus.
The in situ halo stars with high [{Mg}/{Fe}] are likely those heated by this event from the disk present at that time \citep{Helmi2018a,Belokurov2020a,Gallart2019a,DiMatteo2019a}.

In addition to these two major populations, there seems to be an additional, highly retrograde component in the Milky Way halo.
The Gaia data revealed that there is an overdensity of stars in the space of kinematics that has large retrograde motion and large orbital energy \citep[e.g., ][]{Myeong2018c,Koppelman2018a,Yuan2020a,Naidu2020a}, which is now widely called {Sequoia}.
Since Sequoia is less prominent and has a lower mean metallicity than Gaia-Enceladus, this could be a disrupted dwarf galaxy smaller than Gaia-Enceladus \citep{Myeong2019a,Koppelman2019a,Matsuno2019a,Naidu2020a,Feuillet2021a}.
However, the picture is somewhat confusing because the region occupied by Sequoia has been suggested to contain multiple components \citep{Naidu2020a}.
\citet{Helmi2018a} suggest that the progenitor of Gaia-Enceladus can also deposit stars onto Sequoia-like orbits depending on the morphology and the inclination of its initial orbit, which is supported by numerical simulations \citep{Koppelman2019a}. 
Another complication arises in the selection and definition of Sequoia stars; while \citet{Matsuno2019a}, \citet{Koppelman2019a}, \citet{Naidu2020a}, and \citet{Aguado2021a} used selections in the angular momentum ($L_z$) and orbital energy ($E$) of the stars ($L_z-E$ selection), \citet{Myeong2019a}, \citet{Monty2020a}, and \citet{Feuillet2021a} mostly used normalized actions ($\tilde{\textbf{J}}$ selection).
In the latter case, the selected stars extend to a much lower orbital energy \citep[e.g.,][]{Feuillet2021a}.

Chemical abundance analysis from high-resolution spectroscopy is crucial to understand the properties of Sequoia.
Differences in abundance ratios imply different conditions of star formation, for example, star formation with a different efficiency, which, in turn, could be related to the mass of the progenitor galaxy and/or its environment. 
Although chemical abundance ratios have been well studied for Gaia-Enceladus, in situ stars, and surviving dwarf galaxies \citep[e.g.,][]{Venn2004a,Tolstoy2009,Nissen2010,Hasselquist2021a}, the current understanding about the chemical abundance ratios of Sequoia is much less clear.

Long before the discovery of Sequoia as a kinematic overdensity, \citet{Venn2004a} pointed out the systematically low [$\alpha$/{Fe}] ratios of highly retrograde stars; they showed that the very low [$\alpha$/{Fe}] ratios seen among the outermost halo stars in \citet{Stephens2002} are rather related to their large retrograde motion.
For the first time, \citet{Matsuno2019a} have recently indicated the connection between Sequoia and the results from \citet{Venn2004a}.
They selected stars in an overdensity with large retrograde motion at a high orbital energy ($L_z-E$ selection), which later named Sequoia, from the Stellar Abundances for Galactic Archaeology Database \citep{Suda2008,Suda2011,Yamada2013,Suda2017a}.
They show that the selected stars have, on average, lower [{Na}/{Fe}], [{Mg}/{Fe}], and [{Ca}/{Fe}] than Gaia-Enceladus or in situ stars at [{Fe}/{H}]$\sim-1.5$.
This conclusion is supported by \citet{Monty2020a}, who recalibrated the abundances and recalculated the kinematics of stars studied by \citet{Stephens2002}.
Among their Sequoia stars selected with the $\tilde{\textbf{J}}$ selection, stars with a low binding energy, corresponding to a $L_z-E$ selection, show low values of [{Mg}/{Fe}] and [{Ca}/{Fe}], while stars with a higher binding energy have higher Mg and Ca abundances.
Additionally, the proper-motion pair, HD134439 and HD134440, which has large retrograde Galactic motion \citep{King1997a,Lim2021a} and indeed satisfies most of the $L_z-E$ selections, has been known to have low $\alpha$-element abundances \citep{King1997a,Chen2006a,Chen2014a,Reggiani2018a,Lim2021a}.
Among these studies, \citet{Reggiani2018a} have indeed confirmed that the $\alpha$-element abundances of HD134439 and HD134440 are even lower than NS10's low-$\alpha$ halo population.
Data from GALAH DR3 also seem to support the low-$\alpha$ abundance of Sequoia \citep[see $\alpha$-element abundance presented in][who used a $L_z-E$ selection]{Aguado2021a}.
The same feature is, however, not clearly seen in data from APOGEE as presented in \cite{Koppelman2019a} with a $L_z-E$ selection, and \citet{Myeong2019a} and \citet{Feuillet2021a} with a $\tilde{\textbf{J}}$ selection.

\begin{table*}
  \caption{Summary of the data \label{tab:obs}}
  \centering
  \begin{tabular}{lrrrrrr}
\hline\hline
Object & Gaia EDR3 source id & Telescope & Resolution           &  $S/N_1$ & $S/N_2$ &   $S/N_3$   \\\hline
1336\_6432       &  1336408284224866432 & Subaru     &  80,000  &   94     &   139   &   137       \\
2657\_5888       &  2657496656325125888 & Subaru     &  80,000  &   41     &    79   &    92       \\
2813\_6032       &  2813331813720876032 & Subaru     &  80,000  &  130     &   173   &   126       \\
2870\_9072       &  2870313110476579072 & Subaru     &  80,000  &   88     &   118   &    52       \\
3336\_0672       &  3336204190352220672 & Subaru     &  80,000  &   90     &   194   &   114       \\
3341\_2720       &  3341934256545182720 & Subaru     &  80,000  &  126     &   140   &    99       \\
4587\_5616       &  4587905579084735616 & Subaru     &  80,000  &   48     &    81   &    75       \\
4850\_5696 &  4850673911632285696 & Magellan   &  32,000-40,000 &  207     &   128   &   176       \\\hline
G90-36          &   876358870971624320 & Keck       &  48,000   &   41     &    67   &    50       \\
G115-58         &  1011379899590855936 & Subaru     & 100,000   &   90     &   147   &    98       \\
HIP28104        &  2910503176753011840 & VLT        &  50,000   &   93     &   161   &   181       \\
HIP98492        &  4299974407538484096 & Keck       &  72,000   &   77     &   133   &    83       \\\hline
G112-43         &  3085891537839267328 & Subaru     & 100,000   &  214     &   208   &    99       \\
CD$-48^{\circ}$02445      &  5551565291043498496 & VLT        &  50,000   &  130     &   289   &   258       \\
HD59392         &  5586241315104190848 & VLT        &  50,000   &  170     &   441   &   248       \\
\hline
    
  \end{tabular}
\tablefoot{We obtained new high-resolution spectra for the eight Sequoia stars in the first group, and archival high-resolution spectra for the four Sequoia stars in the second group and for the three standard stars in the last group. The $S/N_1$, $S/N_2$, and $S/N_3$ were measured at around $4500,\,5533,\,\mathrm{and}\, 6370\,\mathrm{\AA}$, respectively, and converted to per $0.01\,\mathrm{\AA}$.}
 \end{table*} 

\begin{sidewaystable*}
  \caption{Target information\label{tab:kinematics_photo}}
  \centering
  \begin{tabular}{lrrrrrrrrrr}
\hline\hline
Object & $\pi$ & $\sigma(\pi)$ & $G$ & $B_p-R_p$ & $E(B-V)$ & RV & $L_z$ & $\sigma(L_z)$ & $E$ & $\sigma(E)$ \\
& (mas) & (mas) & & &  & $(\mathrm{km\,s^{-1}})$ & $(\mathrm{kpc\,km\,s^{-1}})$ & $(\mathrm{kpc\,km\,s^{-1}})$ & $(\times 10^5\,\mathrm{km^2\,s^{-2}})$ & $(\times 10^5\,\mathrm{km^2\,s^{-2}})$ \\\hline
1336\_6432       &      2.770 &      0.011 & 11.735 &  0.626 &  0.038 & -560.2 &      -3000 &         12 &     -1.022 &      0.005 \\
2657\_5888       &      1.798 &      0.024 & 13.960 &  0.924 &  0.024 & -348.4 &      -2934 &         52 &     -0.936 &      0.031 \\
2813\_6032       &      2.929 &      0.020 & 11.922 &  0.707 &  0.098 & -461.9 &      -1787 &         12 &     -1.218 &      0.005 \\
2870\_9072       &      1.397 &      0.017 & 12.885 &  0.855 &  0.116 & -392.1 &      -1628 &         25 &     -1.258 &      0.019 \\
3336\_0672       &      5.875 &      0.030 & 11.603 &  0.794 &  0.000 &  -13.9 &      -1722 &         26 &     -1.251 &      0.011 \\
3341\_2720       &      1.344 &      0.014 & 13.281 &  0.792 &  0.130 & -137.3 &      -2478 &         92 &     -0.893 &      0.038 \\
4587\_5616       &      3.675 &      0.019 & 12.771 &  0.929 &  0.010 & -600.2 &      -3147 &          9 &     -1.016 &      0.004 \\
4850\_5696       &      3.981 &      0.014 & 10.987 &  0.616 &    0.0 &  430.4 &      -1604 &         11 &     -1.299 &      0.003 \\
G90-36          &      4.120 &      0.013 & 12.559 &  0.828 &  0.004 &  267.4 &      -2880 &         46 &     -1.004 &      0.019 \\
G115-58         &      2.224 &      0.019 & 11.973 &  0.657 &  0.036 &  226.2 &      -1799 &         62 &     -1.246 &      0.017 \\
HIP28104        &      2.590 &      0.011 & 12.080 &  0.620 &  0.014 &  253.6 &      -2504 &         30 &     -1.110 &      0.019 \\
HIP98492        &      2.660 &      0.018 & 11.373 &  0.898 &  0.076 & -266.4 &      -1791 &         36 &     -1.229 &      0.028 \\\hline
G112-43         &      5.595 &      0.019 & 10.063 &  0.686 &  0.000 &        &            &            &            &            \\
CD$-48^{\circ}$02445      &      5.369 &      0.014 & 10.418 &  0.616 &    0.0 &        &            &            &            &            \\
HD59392         &      6.382 &      0.013 &  9.576 &  0.685 &    0.0 &        &            &            &            &            \\
\hline
  \end{tabular}
   \tablefoot{Parallax and photometric information from Gaia EDR3. The extinction coefficient is from \citet{Green2019a}, except for 4850\_5696, CD$-48^{\circ}$02445, and HD59392, for which $E(B-V)=0.0$ is assumed. Radial velocity is from our newly obtained spectra, except for G90-36, HIP28104 \citep{Stephens2002}, G115-58 \citep{Ishigaki2012}, and HIP98492 \citep{Omalley2017a}.}
\end{sidewaystable*}

In summary, several studies seem to have shown that Sequoia stars have lower $\alpha$-element abundances than Gaia-Enceladus on average if they are selected according to $L_z$ and $E$.
However, the magnitude of the difference is comparable to the typical uncertainties of abundance ratios for individual stars. 
This small difference together with the existence of multiple ways to kinematically select Sequoia stars have hampered a clear understanding.
High-precision chemical abundance from a differential abundance analysis on high signal-to-noise ($S/N$) spectra would allow us to robustly detect the abundance difference, if any, as shown by \citet{Nissen2010,Nissen2011} to be powerful to characterize the abundance difference between accreted and in situ halo stars.

The aim of this study is to carry out a high-precision abundance analysis for Sequoia stars selected by the $L_z-E$ selection using both newly obtained high-resolution spectra and archival spectra, and to compare the results with other halo stars.
The comparison halo stars come from NS10 at $-1.5\lesssim [\mathrm{Fe/H}]\lesssim -0.7$ and \citet[][R17]{Reggiani2017a}, who carried out a high-precision abundance analysis for stars with $-2.5\lesssim[\mathrm{Fe/H}]\lesssim-1.5$. 
As we see below, we clearly detected very low-$\alpha$ element abundances for Sequoia stars.
We describe our target selection and the data in Section~\ref{sec:obs} and the abundance analysis in Section~\ref{sec:analysis}.
After briefly introducing our results in Section~\ref{sec:result}, we provide a discussion in Section~\ref{sec:discussion} and present our conclusions in Section~\ref{sec:conclusion}. 

\begin{figure*}
\centering
\includegraphics[width=1.0\textwidth]{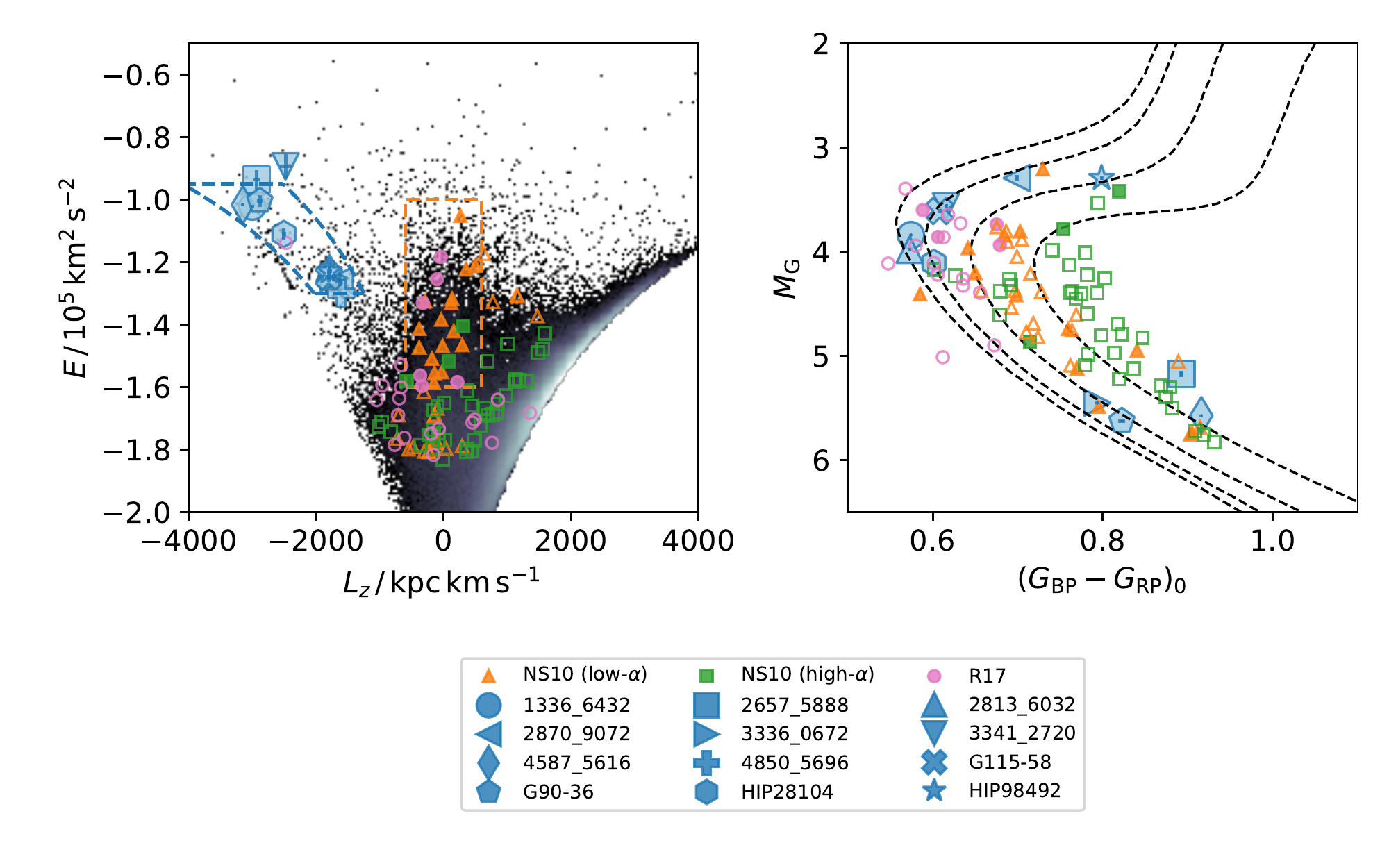}
\caption{Kinematic and photometric properties of the stars. Left: Angular momentum and orbital energy of the observed stars and the comparison stars. We also show the distribution of all stars in Gaia EDR3 with good astrometry (a relative parallax uncertainty smaller than 20\%) and Gaia DR2 radial velocity. The blue dotted lines represent the Sequoia selection from \citet{Koppelman2019a}. Stars in the orange box are used to define the chemical abundance trends of Gaia-Enceladus. NS10 and R17 stars within this box are shown with filled symbols, while those outside of it are shown with open symbols. Right: Gaia EDR3 color-magnitude diagram of the program stars. We also plotted four PARSEC isochrones with the age of $12\ \mathrm{Gyr}$ and [{Fe}/{H}]$=-2.0,\ -1.5,\ -1.0,{\rm and }-0.5$ (from left right). We note that only stars with available extinction estimates from \citet{Green2019a} are plotted for the NS10 and R17 samples. \label{fig:kinematicsCMD}}
\end{figure*}

\section{Observations and target selection\label{sec:obs}}

We obtained new high-$S/N$, high-resolution spectra for nine Sequoia member candidates with the Subaru Telescope (for eight stars) and with Magellan (for one star). 
Out of the nine stars, one star (Gaia EDR3 360456543361799808) turned out to have a very different radial velocity from the value used for the selection (see Appendix \ref{appendixA}), and hence it can no longer be regarded as a part of Sequoia.
We also collected high-$S/N$ archival high-resolution spectra for four Sequoia candidates and for three standard stars from Subaru, Keck, and Very Large Telescopes (VLT).
The data and target information are summarized in Table~\ref{tab:obs} and Table~\ref{tab:kinematics_photo}, respectively.

The Subaru observations were conducted with the High Dispersion Spectrograph \citep[HDS;][]{Noguchi2002} from November 8--10, 2019. 
We used the standard setup StdYd of HDS, which provides a wavelength coverage of $4000-6800\ \mathrm{\AA}$, and the image slicer \#2 \citep{Tajitsu2012a}, which yields $R\sim 80,000$.
Two to eight exposures were taken for each object and the total exposure time ranged from 20 minutes to 4 hours depending on the brightness of the stars.
We reduced the data using an \texttt{IRAF}\footnote{IRAF is distributed by the National Optical Astronomy Observatory, which is operated by the Association of Universities for Research in Astronomy (AURA) under a cooperative agreement with the National Science Foundation} script, \texttt{hdsql}\footnote{\url{http://www.subarutelescope.org/Observing/Instruments/HDS/hdsql-e.html}}, which includes a CCD linearity correction, scattered light subtraction, aperture extraction, flat-fielding, wavelength calibration, and a heliocentric velocity correction.

The Magellan observation was conducted with the Magellan Inamori Kyocera Echelle \citep[MIKE;][]{Bernstein2003a} on December 29, 2019.
Although the MIKE spectrum has a wide spectral coverage from $3350\ \mathrm{\AA}$ to $9300\ \mathrm{\AA}$, we only used $4000-6800\ \mathrm{\AA}$ to maintain consistency with the HDS spectra.
The slit width was $0.70''$, which yields $R\sim 40,000$ for the region bluer than $4950\,\mathrm{\AA}$ and $R\sim 32,000$ for the redder part.

We searched reduced archive spectra using the Japanese Virtual Observatory portal for archive HDS spectra\footnote{\url{http://jvo.nao.ac.jp/portal/subaru/hds.do}}, the Keck Observatory Archive (KOA) for data taken with the High Resolution Echelle Spectrometer \citep[HIRES;][]{Vogt1994} on Keck, and European Southern Observatory Science Archive Facility for data taken with the Ultraviolet and Visual Echelle Spectrograph \citep[UVES;][]{Dekker2000a}. 
In the present study, we used archive spectra for four Sequoia stars and three standard stars (Table~\ref{tab:obs}).

All the spectra were normalized by fitting continua with cubic splines after combining multiple exposures for individual objects.
For objects for which we conducted new observations, the radial velocity was measured by comparing the observed wavelengths of \ion{Fe}{I} absorption lines with the values measured by laboratory experiments.
We adopted literature measurements of radial velocities for stars for which spectra were taken from archives (Table~\ref{tab:kinematics_photo}).

All the candidate Sequoia member stars were selected based on their angular momentum around the $z$-axis of the Milky Way ($L_z$) and orbital energy ($E$).
We combined Gaia DR2 astrometry with radial velocity measurements provided in Gaia DR2 and LAMOST DR4 for the target selection.
Table~\ref{tab:kinematics_photo} and Figure~\ref{fig:kinematicsCMD} include updated kinematic information for the program stars, where astrometry is now taken from Gaia EDR3 and the radial velocity comes from high-resolution spectroscopy.
Although we updated the astrometry and radial velocity, the orbital parameters obtained did not change significantly.
Here we assumed that the Sun is located at $R_0=8.21\ \mathrm{kpc}$ \citep{McMillan2017a} and $z_0=20.8\ \mathrm{pc}$ \citep{Bennett2019a} and moving at $11.1\ \mathrm{km\ s^{-1}}$ toward the Galactic anticenter \citep{Schoenrich2010a}, $245.3\ \mathrm{km\ s^{-1}}$ in the Galactic rotation direction \citep{Reid2004a,McMillan2017a}, and $7.25\ \mathrm{km\ s^{-1}}$ toward the Galactic north pole \citep{Schoenrich2010a}.
The Milky Way potential used is from \citet{McMillan2017a}.
The calculation of the orbital energy was conducted with the software \textsc{Agama} \citep{Vasiliev2019a}.
We estimated uncertainties through a Monte Carlo method.
All the Sequoia candidates have $L_z<-1600\ \mathrm{kpc\ km\ s^{-1}}$ and $E>-1.3\times 10^5\ \mathrm{km^2\ s^{-2}}$.
We also removed stars with $[\mathrm{Fe/H}]>-1.0$ if they had metallicity estimates from LAMOST.

We note that our kinematic selections are based on our knowledge from Gaia~DR2.\ In addition, updated selections will be available with more recent and future Gaia data releases \citep[see][]{Loevdal2022a,RuizLara2022a}.

To suppress the effect of systematic uncertainties that depend on stellar spectral types and to put our derived abundances onto the same scale as previous studies, we limited the sample to stars around the main-sequence turn-off region using the Gaia DR2 color-magnitude diagram.
The updated photometric information with Gaia EDR3 is summarized in Table~\ref{tab:kinematics_photo} and Figure~\ref{fig:kinematicsCMD}.
The extinction was estimated using the three-dimensional dust map by \citet{Green2019a} and the extinction coefficients were estimated following \citet{Casagrande2021a}.
For three objects, 4850\_5696, HD59392, and CD$-48^{\circ}$02445 not in the coverage of \citet{Green2019a}, we assumed their extinctions to be negligible since \citet{RuizDern2018a} and \citet{Lallement2019a} provide very small estimates for them ($E(B-V)<0.01$). 
We note that two stars (2657\_5888 and 4587\_5616) lie along metal-rich isochrones.
We discuss possible origins of the offset and effects on derived abundances in Appendix~\ref{sec:appendixC}.

We note that two additional stars (BD+09 2190 and HE1509-0252) were included in the list of Sequoia candidates with archive spectra, but they are not analyzed in the present study.
These two objects have a much lower metallicity ([{Fe}/{H}]$=-2.63$ \citep{Ishigaki2012} and $-2.85$ \citep{Cohen2013a}, respectively) than the rest of the sample, and hence they are not suitable for the differential abundance analysis conducted in this study.
The removal of these two stars and the metallicity selection ($[\mathrm{Fe/H}]<-1.0$) for LAMOST stars result in narrower metallicity dispersion among our sample (0.19 dex) compared to $\sim 0.3\,\mathrm{dex}$ reported for Sequoia in the literature \citep[e.g.,][]{Matsuno2019a}.

To compare the properties of Sequoia stars with those of in situ and Gaia-Enceladus stars, we contrast chemical abundances of our Sequoia candidates with stars studied by NS10 and R17. 
These stars are also plotted in Figure~\ref{fig:kinematicsCMD} and cover a similar region of the color magnitude diagram. 
The orbital parameters of most of the NS10 stars and R17 stars are clearly different from the region of Sequoia.
We note that the most retrograde star in the R17 sample is HIP28104, which is regarded as a Sequoia member candidate and is included in our analysis.
The abundance of this star reported by R17 is marked in all our abundance figures with a special symbol.

\section{Abundance analysis\label{sec:analysis}}

We derived stellar parameters and chemical abundances based on a differential abundance analysis adopting HD59392 as the reference star.
Together with the high quality of the data, this approach enabled us to achieve high precision abundance measurements.
In this section, we describe the analysis method, validate the results by comparing them with the literature, and homogenize abundances from the present study, NS10, and R17.
For the abundance analysis, we used the November 2019 version of MOOG \citep{Sneden1973} through a python wrapper, $\texttt{q}^2$\citep{Ramirez2014}, and adopted the standard \texttt{MARCS} model atmospheres \citep{Gustafsson2008}. 

\subsection{Equivalent widths measurements}

\begin{table*}
  \caption{Linelist and line-by-line abundance\label{tab:linelist}}
  \centering
  \begin{tabular}{l*{7}{r}}\hline\hline
Object & species & $\lambda$ & $\chi$ & $\log gf$ & $EW$ & $\sigma(EW)$  &    $A(X)$ \\
       &         & ($\mathrm{\AA}$) & ($\mathrm{eV}$) &  & ($\mathrm{m\AA}$) & ($\mathrm{m\AA}$) & (dex) \\
\hline
1336\_6432   &NaI  &  5889.959&     0.000&    -0.193&     133.6&       6.2&   4.230\\   
1336\_6432   &NaI  &  5895.910&     0.000&    -0.575&     107.0&       5.0&   4.108\\   
1336\_6432   &MgI  &  4167.271&     4.346&    -0.746&      43.7&       2.1&   5.969\\   
1336\_6432   &MgI  &  5167.321&     2.709&    -0.854&     125.8&       5.8&   6.125\\   
1336\_6432   &MgI  &  5172.684&     2.712&    -0.363&     173.0&       8.0&   6.118\\ \hline\hline  
 $\sigma(A)_{T_{\rm eff}}$ & $\sigma(A)_{\log g}$ &  $\sigma(A)_{v_t}$ &  $\sigma(A)_{[\mathrm{Fe/H}]}$ & $\sigma(A)_{EW}$ & $s_X$ & weight \\
(dex) & (dex) & (dex) & (dex) & (dex) & (dex) & \\ \hline
    0.053&    -0.009&    -0.030&    -0.003&     0.108&     0.000&    60.883\\
    0.034&    -0.003&    -0.024&     0.007&     0.099&     0.000&    78.054\\
    0.026&    -0.003&    -0.003&     0.001&     0.053&     0.000&   204.968\\
    0.055&    -0.012&    -0.023&     0.001&     0.096&     0.000&    21.625\\
    0.064&    -0.017&    -0.014&     0.001&     0.080&     0.000&    10.299\\

\hline
  \end{tabular}
\tablefoot{The full table is available online at CDS and a portion of the table is shown here.}
 \end{table*} 

\begin{figure}
\centering
\includegraphics[width=0.5\textwidth]{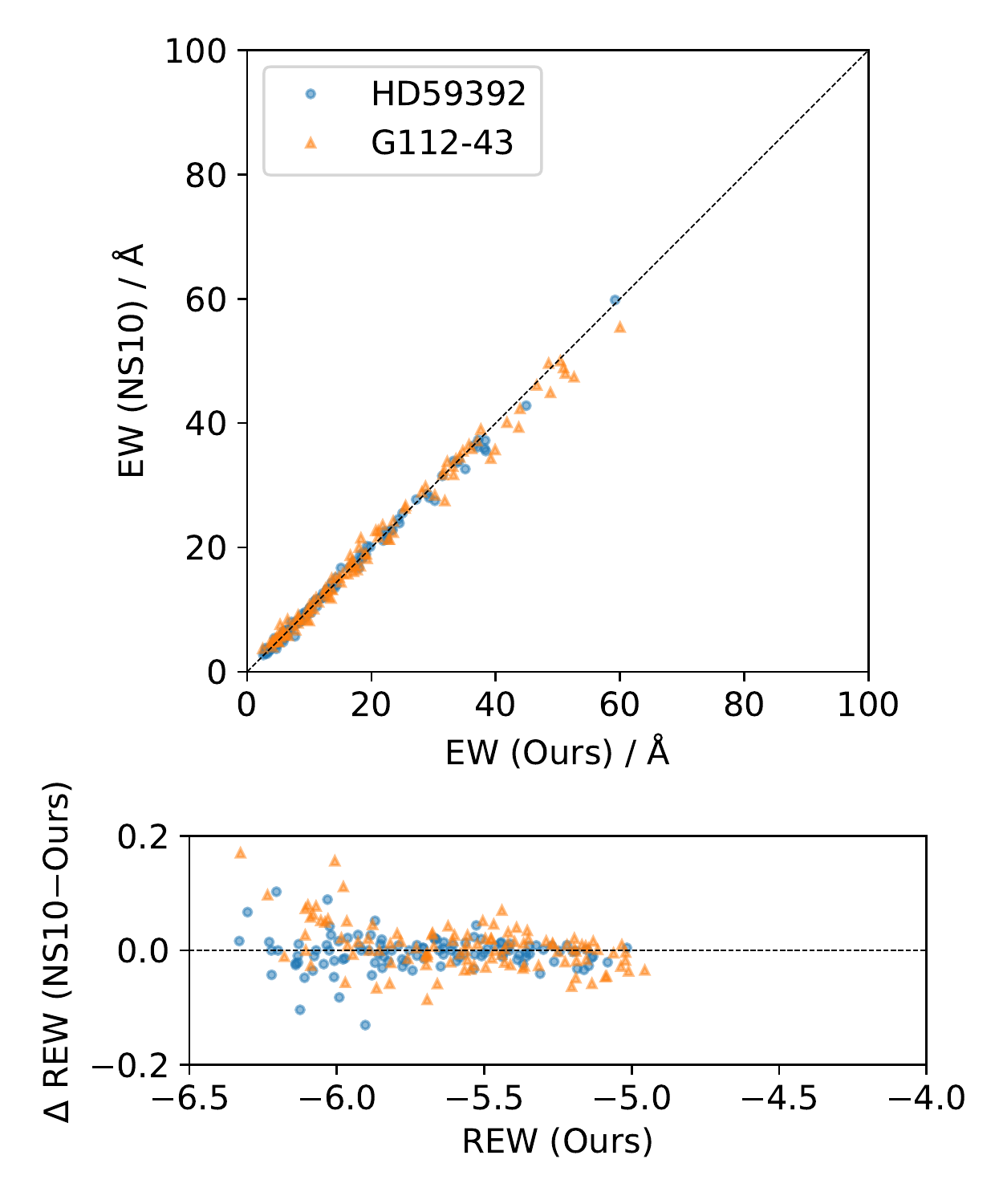}
\caption{Equivalent width comparison with NS10.\label{fig:ewNS10}}
\end{figure}

\begin{figure}
\centering
\includegraphics[width=0.5\textwidth]{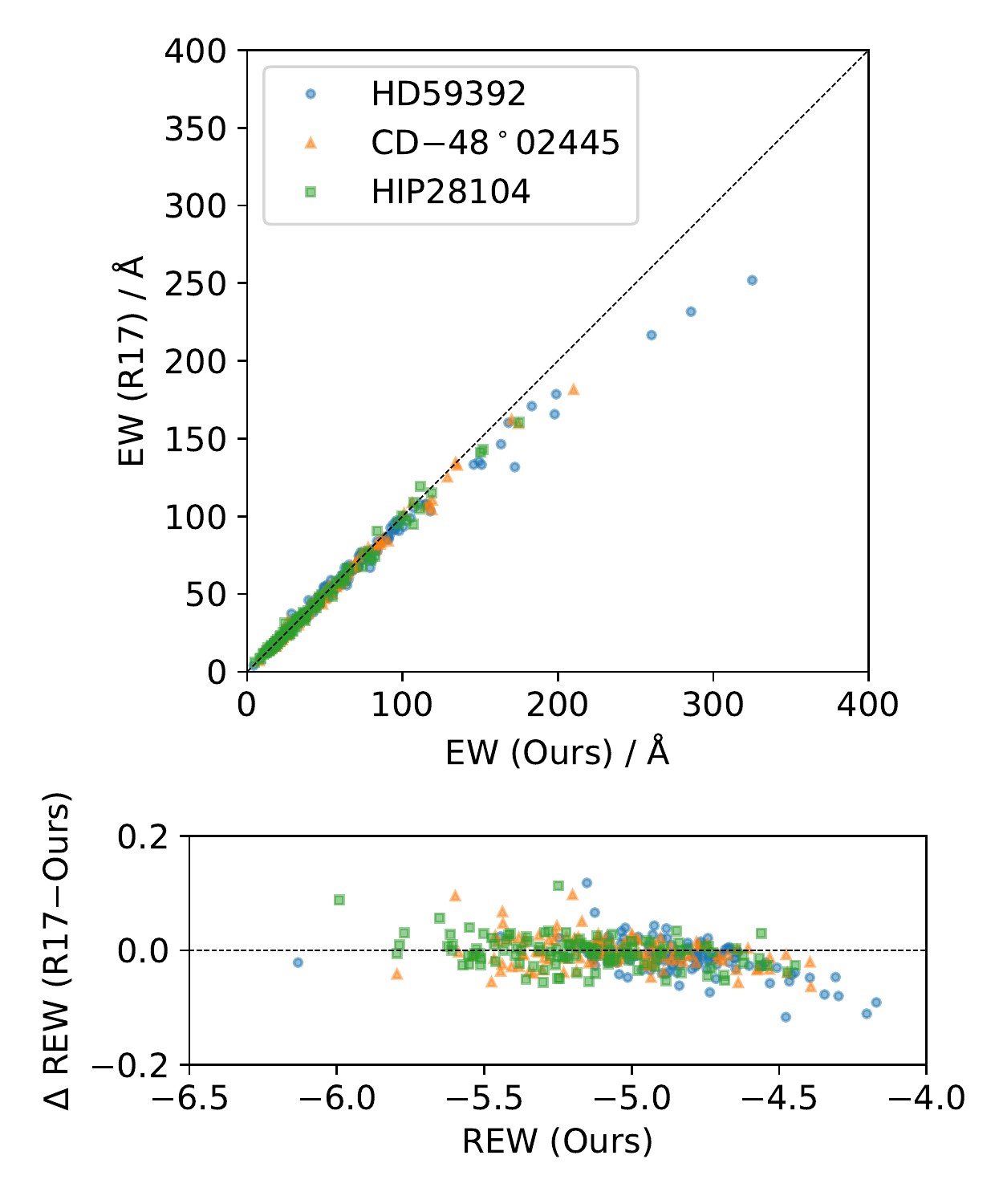}
\caption{Equivalent width comparison with R17. We note that this figure has a wider range in equivalent widths compared to Figure~\ref{fig:ewNS10}. \label{fig:ewR17}}
\end{figure}

Table~\ref{tab:linelist} provides the line list and measured equivalent widths ($EW$).
The line selection follows NS10 and R17, but the $\log gf$ values were updated for homogenization purposes. 
We measured equivalent widths of the lines by fitting Voigt profiles, of which the results were visually inspected.
Stellar parameter determination and subsequent elemental abundance measurements are based on these measured equivalent widths unless otherwise stated.
Spectral synthesis was applied to Si, Mn, Zn, and Y.
The equivalent widths of the lines from these elements were not measured by the Voigt profile fitting, but they are estimates based on a synthetic spectrum. 
Hyperfine structure splitting was taken into consideration for Na, Mn, and Ba.

We have two and three stars in common with NS10 and R17, respectively.
The equivalent widths measured from archive spectra are compared with the values reported in the literature for these objects in Figures~\ref{fig:ewNS10} and \ref{fig:ewR17}\footnote{The equivalent widths for the objects in R17 were kindly provided by H. Reggiani in private communication.}.
In comparison with NS10, G112-43 particularly offers an opportunity to confirm that different telescopes yield consistent spectra because we used a Subaru/HDS archive spectrum for this object (Table~\ref{tab:obs}), while NS10 used one from VLT/UVES.
We find excellent agreement in the comparison with NS10 for HD59392 and for G112-43. 
The average differences in reduced equivalent widths ($REW=\log(EW/\lambda)$) are $\Delta REW=-0.004$ ($\sigma=0.029$) and $\Delta REW= 0.007$ ($\sigma=0.041$), respectively.

The agreements with R17 are also good, but our measured equivalent widths are larger than theirs for large equivalent widths, which is likely due to the different assumption for the line shape (Voigt and Gaussian profiles).
We confirmed that Voigt profiles provide better fits for the strongest lines than Gaussian profiles through visual inspection.
Despite the offset at large equivalent widths, the average differences in reduced equivalent widths are small for all three objects ($\Delta REW=-0.012$, $\sigma=0.030$ for HD59392; $\Delta REW=-0.001$, $\sigma=0.026$ for CD$-48^{\circ}$02445; and $\Delta REW=-0.001$, $\sigma=0.029$ for HIP28104).

We estimated the uncertainties in the equivalent widths using the formula provided by \citet{Cayrel1988a}. 
Although the formula predicts very small fractional uncertainties for strong lines, Figures~\ref{fig:ewNS10} and \ref{fig:ewR17} both show that this is not the case.
Since this holds true even for the strongest lines, this is not likely dependent on $S/N$.
Therefore, based on the scatter in $\Delta REW$ at $-5.5<REW<-4.5$, which is $0.015-0.025\,\mathrm{dex}$, we quadratically added $0.02\,\mathrm{dex}$, which is equivalent to 4.6\%, as an error floor to the uncertainties on the equivalent widths. 

\subsection{Stellar parameter determination\label{sec:parameters}}

We determined the stellar parameters by combining an analysis of iron lines, as well as photometric and astrometric information from the Gaia mission.
We estimated the effective temperature ($T_{\rm eff}$) and microturbulent velocity ($v_t$) by minimizing the correlation coefficients between iron abundances derived from individual neutral iron lines, and excitation potentials and $REW$, respectively.

We calculated the surface gravity ($\log g$) from 
\begin{eqnarray}
\log g &=& \log g_{\odot} + \log(M/M_\odot) - 2\log(R/R_{\odot})\nonumber\\
&=&\log g_{\odot} + \log(M/M_\odot) + 4\log(T_{\rm eff}/T_{\rm eff,\odot}) - \log (L/L_\odot),\label{eq:logg}
\end{eqnarray}
where $M$, $R$, and $L$ are the mass, radius, and luminosity of the star, respectively.
The mass was obtained by finding the stellar model that describes the position of the star in the color-absolute magnitude diagram best.
Three parameters, age ($\tau$), initial mass ($M_{\rm ini}$), and $\mathrm{[Fe/H]_{\rm model}}$, were varied to find the best model.
We maximized $p(\theta|\bf x)$ through an ensemble Markov chain Monte Carlo sampling using \texttt{emcee} \citep{emcee}, where $\theta$ is $(\tau,\,M_{\rm ini},\,\mathrm{[Fe/H]_{\rm model}})$, $\bf x$ is $(M_G,(Bp-Rp)_0,\mathrm{[Fe/H]}_{\rm input})$, and $p(\theta|\textbf{x})\propto L(\textbf{x}|\theta)p(\theta)$.
The likelihood $L(\textbf{x}|\theta)$ was expressed as a multivariate Gaussian distribution $\mathcal{N}(\textbf{x}_{\rm model}|\textbf{x}_{\rm obs},\Sigma)$, in which $\Sigma$ reflects the observational uncertainties.
For the uncertainty of $M_G$ and $(Bp-Rp)_0$ and the covariance between the two, we considered the uncertainty in Gaia photometry and parallax, and extinction.
We adopted a $0.2\,\mathrm{dex}$ uncertainty for $[\mathrm{Fe/H}]_{\rm sp}$ to take systematic uncertainties into account.
We used the initial mass function of \citet{Kroupa2003a} as the prior for the $M_{\rm ini}$ and a flat prior for the age between 1 and 20 Gyr.
The luminosity was obtained from $M_G$ and the bolometric correction by \citet{Casagrande2018a}.

First estimates of the uncertainties in the stellar parameters were obtained in the following way:
the uncertainties in $T_{\rm eff}$ and $v_t$ were estimated by finding the ranges of the values that provide the corresponding correlation coefficients consistent with zero at the $1\sigma$ level;
the $\log g$ uncertainty was computed by randomly sampling $M$ and $L$ in Eq.~\ref{eq:logg} following their uncertainties, where covariances between $M$, $T_{\rm eff}$, and $L$ were assumed to be negligible;
and the uncertainty of $[\mathrm{Fe/H}]_{\rm sp}$ was obtained from the standard deviation of the iron abundances from individual lines divided by the square root of the number of lines used.
These estimates, however, do not take correlations between parameters into consideration.
For example, since $[\mathrm{Fe/H}]_{\rm sp}$ clearly depends on the assumed values of the other parameters, we need to propagate the uncertainties in the other parameters into the uncertainty estimate for $[\mathrm{Fe/H}]_{\rm sp}$.
We corrected the estimated uncertainties by properly considering the correlations between parameters following the method described in Appendix~\ref{app:uncertainty}.
As a result of this procedure, we also obtained covariances between the estimated parameters.

Since we adopted a differential abundance analysis, the parameters of the standard star HD59392 determine the scale of our parameters.
We adopted $T_{\rm eff}=6012\,\mathrm{K}$ and $[\mathrm{Fe/H}]_{\rm sp}=-1.6$ (NS10), and we re-determined $\log g$ using the astrometric information.
The microturbulence was updated to $v_t=1.4\,\mathrm{km\,s^{-1}}$ so that the neutral iron abundances derived from individual lines do not depend on the line strength.

The stellar parameters and their uncertainties obtained as just described are provided in Table~\ref{tab:parameters}.
We note that the $[\mathrm{Fe/H}]_{\rm sp}$ values in Table~\ref{tab:parameters} are not the same as $[\mathrm{Fe/H}]_{\mathrm{I}}$ or $[\mathrm{Fe/H}]_{\mathrm{II}}$ which we report in the next section. 
This is because the computation of weights we use in the next section for abundance and its uncertainty estimates requires predetermined stellar parameters and their uncertainties.
Since the stellar parameter determination process itself needs an iterative process, here, we adopted the simple mean of the abundances from individual lines to simplify the computation while we consider a weighted average in the next section.

\begin{sidewaystable*}
  \caption{Stellar parameters\label{tab:parameters}}
  \centering
  \begin{tabular}{l*{14}{r}}
\hline\hline
Object & $T_{\rm eff}$ & $\sigma(T_{\rm eff})$ &  $\log g$ & $\sigma(\log g)$ & $v_t$ & $\sigma(v_t)$ &  $[\mathrm{Fe/H}]_{\rm sp}$ & $\sigma([\mathrm{Fe/H}]_{\rm sp})$ & $\rho_{T_{\rm eff},\log g}$ & $\rho_{T_{\rm eff},v_t}$ & $\rho_{T_{\rm eff},[\mathrm{Fe/H}]_{\rm sp}}$ &  $\rho_{\log g,v_t}$ & $\rho_{\log g,[\mathrm{Fe/H}]_{\rm sp}}$ & $\rho_{v_t,[\mathrm{Fe/H}]_{\rm sp}}$  \\
&($\mathrm{K}$)&($\mathrm{K}$)&(dex) &(dex) &($\mathrm{km\,s^{-1}}$)&($\mathrm{km\,s^{-1}}$)&(dex)&(dex)& & & & & & \\\hline
1336\_6432       & 6475 &   65 &  4.166&  0.034&  1.512&  0.102& -1.691&  0.030&  0.485&  0.269&  0.280&  0.108&  0.448& -0.540 \\
2657\_5888       & 5317 &  152 &  4.314&  0.056&  0.784&  0.530& -1.535&  0.053&  0.845&  0.925& -0.751&  0.749& -0.504& -0.807 \\
2813\_6032       & 6414 &   63 &  4.209&  0.035&  1.624&  0.102& -1.662&  0.029&  0.425&  0.435&  0.206&  0.071&  0.468& -0.264 \\
2870\_9072       & 5815 &   60 &  3.788&  0.046&  1.272&  0.066& -1.467&  0.023&  0.351&  0.704&  0.272&  0.123&  0.743&  0.035 \\
3336\_0672       & 5575 &  148 &  4.471&  0.051&  1.062&  0.427& -1.691&  0.033&  0.816&  0.951& -0.542&  0.738& -0.228& -0.625 \\
3341\_2720       & 6538 &   78 &  4.082&  0.038&  1.456&  0.124& -1.831&  0.031&  0.537&  0.355&  0.355&  0.169&  0.525& -0.425 \\
4587\_5616       & 5481 &   80 &  4.496&  0.033&  1.329&  0.235& -1.768&  0.048&  0.661&  0.688& -0.304&  0.343&  0.009& -0.470 \\
4850\_5696       & 6448 &   65 &  4.194&  0.033&  1.415&  0.107& -1.728&  0.027&  0.455&  0.665&  0.018&  0.265&  0.357& -0.287 \\
G90-36          & 5394 &   46 &  4.476&  0.030&  1.210&  0.172& -1.670&  0.045&  0.262&  0.553& -0.306& -0.069&  0.236& -0.491 \\
G115-58         & 6187 &   69 &  4.012&  0.041&  1.361&  0.069& -1.394&  0.021&  0.309&  0.670&  0.126& -0.023&  0.770& -0.249 \\
HIP28104        & 6468 &   45 &  4.247&  0.042&  1.339&  0.090& -1.986&  0.027&  0.255&  0.507&  0.143&  0.073&  0.539& -0.126 \\
HIP98492        & 5510 &   40 &  3.726&  0.037&  1.130&  0.062& -1.272&  0.021&  0.199&  0.526& -0.171& -0.170&  0.713& -0.358 \\\hline
G112-43         & 6125 &   54 &  4.086&  0.031&  1.412&  0.070& -1.286&  0.019&  0.421&  0.695&  0.137&  0.117&  0.626& -0.214 \\
CD$-48^{\circ}$02445      & 6446 &   58 &  4.205&  0.036&  1.457&  0.111& -1.852&  0.025&  0.421&  0.695&  0.005&  0.244&  0.425& -0.310 \\
HD59392         & 6012 &      &  3.954&       &  1.400&       & -1.600&       &       &       &       &       &       &        \\
\hline
\end{tabular}
\tablefoot{
The last three stars are standard stars and not part of Sequoia.
}
\end{sidewaystable*}

\begin{table}
\caption{Abundances of standard stars \label{tab:abundance_standard}}
\centering
\begin{tabular}{l*{5}{r}}
\hline\hline
     &                     \multicolumn{5}{c}{G112-43   }\\\cline{2-6}
     & $N$ & [{X}/{H}] & $\sigma$ & [{X}/{Fe}] & $\sigma$\\\hline
FeI  &    100&    -1.254&     0.022&       ...&       ...\\
FeII &     10&    -1.256&     0.019&       ...&       ...\\
NaI  &      3&    -1.423&     0.076&    -0.169&     0.074\\
MgI  &      5&    -1.075&     0.038&     0.179&     0.036\\
SiI  &      5&    -1.035&     0.025&     0.219&     0.027\\
CaI  &     16&    -0.977&     0.028&     0.278&     0.025\\
TiI  &     11&    -0.951&     0.044&     0.303&     0.037\\
TiII &      7&    -0.914&     0.035&     0.343&     0.033\\
CrI  &      6&    -1.200&     0.058&     0.055&     0.052\\
MnI  &      4&    -1.527&     0.071&    -0.273&     0.069\\
NiI  &     20&    -1.185&     0.029&     0.069&     0.025\\
ZnI  &      2&    -0.940&     0.036&     0.315&     0.035\\
YII  &      2&    -1.390&     0.037&    -0.134&     0.035\\
BaII &      3&    -1.584&     0.044&    -0.328&     0.041\\\hline\hline
     &                     \multicolumn{5}{c}{CD$-48^{\circ}$02445}\\\cline{2-6}
     & $N$ & [{X}/{H}] & $\sigma$ & [{X}/{Fe}] & $\sigma$\\\hline
FeI  &    111&    -1.814&     0.025&       ...&       ...\\
FeII &     11&    -1.829&     0.021&       ...&       ...\\
NaI  &      2&    -2.091&     0.082&    -0.276&     0.079\\
MgI  &      6&    -1.563&     0.035&     0.251&     0.033\\
SiI  &      2&    -1.414&     0.111&     0.400&     0.112\\
CaI  &     23&    -1.461&     0.026&     0.353&     0.025\\
TiI  &     10&    -1.372&     0.051&     0.442&     0.043\\
TiII &     14&    -1.540&     0.032&     0.289&     0.031\\
CrI  &      5&    -1.786&     0.049&     0.028&     0.041\\
MnI  &      7&    -2.161&     0.048&    -0.346&     0.042\\
NiI  &     13&    -1.844&     0.044&    -0.029&     0.040\\
ZnI  &      1&    -1.657&     0.045&     0.158&     0.043\\
YII  &      2&    -1.749&     0.043&     0.080&     0.040\\
BaII &      4&    -1.985&     0.044&    -0.156&     0.041\\\hline\hline
     &                     \multicolumn{5}{c}{HD59392   }\\\cline{2-6}
     & $N$ & [{X}/{H}] & $\sigma$ & [{X}/{Fe}] & $\sigma$\\\hline
FeI  &    134&   (-1.600)&      ...&  (   ...)&       ...\\
FeII &     20&   (-1.600)&      ...&  (   ...)&       ...\\
NaI  &      4&   (-1.750)&      ...&  (-0.150)&       ...\\
MgI  &      7&   (-1.320)&      ...&  ( 0.280)&       ...\\
SiI  &      8&   (-1.330)&      ...&  ( 0.270)&       ...\\
CaI  &     23&   (-1.220)&      ...&  ( 0.380)&       ...\\
TiI  &     16&   (-1.270)&      ...&  ( 0.330)&       ...\\
TiII &     14&   (-1.270)&      ...&  ( 0.330)&       ...\\
CrI  &      6&   (-1.600)&      ...&  ( 0.000)&       ...\\
MnI  &      8&   (-1.960)&      ...&  (-0.360)&       ...\\
NiI  &     26&   (-1.610)&      ...&  (-0.010)&       ...\\
ZnI  &      2&   (-1.460)&      ...&  ( 0.140)&       ...\\
YII  &      2&   (-1.470)&      ...&  ( 0.130)&       ...\\
BaII &      4&   (-1.670)&      ...&  (-0.070)&       ...\\
\hline
\end{tabular}
\tablefoot{The abundance of HD59392 is identical to NS10 since this is the reference star in our analysis.}
\end{table}

\subsection{Elemental abundances\label{sec:abundance}}

Elemental abundances were obtained through a differential abundance analysis, assuming a one-dimensional plane-parallel atmosphere and local thermodynamic equilibrium (1D/LTE), except for Na. 
Since a Na abundance was sometimes derived from the Na~D lines, the deviation from LTE can be significant. 
We corrected for this effect using the grid provided by \citet{Lind2011a}, which is available through the INSPECT database\footnote{
Data obtained from the INSPECT database, version 1.0 (\url{www.inspect-stars.net})}.

For each species, abundances from individual lines were combined to obtain the best estimate for the abundance of the species following the prescription by \citet{Ji2020a}. 
In short, the final abundance is a weighted mean of the abundances from individual lines.
The weight for a line reflects the uncertainty in its equivalent width and the sensitivity of the abundance to stellar parameters.
An error floor ($s_\mathrm{X}$) was added for individual lines so that the log likelihood,
\begin{align}
\log{\cal L}=&-\frac{1}{2}\sum\frac{(A_i - \bar{A})^2}{\sigma^2(A_i)_{EW}+s_X^2} \\ 
&-\frac{1}{2}\sum\log(\sigma^2(A_i)_{EW}+s_X^2)+\mathrm{constants},
\end{align}
where $A_i$ and $\bar{A}$ are the abundance derived from individual lines and the best estimate of the abundance, respectively, is maximized.
The abundance ratios [{X}/{Y}] were computed with the correlation between the two elemental abundances taken into account.
When computing [{X}/{Fe}], we used the iron abundance from the same ionization state as the species X.
We indicate $1\sigma$ confidence error ellipses for Sequoia stars in the abundance figures to visualize the uncertainties and the covariance of the measured abundances.
The information on abundances derived from individual lines, their sensitivities to the uncertainties in stellar parameters and equivalent widths, weights, and error floors are included in Table~\ref{tab:linelist}.

As stated earlier, the abundances from our analysis, and those from NS10, \citet{Nissen2011}, and R17 were put into the same scale.
Our abundances are anchored to NS10 using the abundance of HD59392.
Although HD59392 was also analyzed by R17, it is located at the high metallicity end of their sample, and hence has large measurement uncertainties in their study.
Therefore, we used CD$-48^{\circ}$02445, which is one of the standard stars used by R17, to move the abundances of R17 into the same scale as ours.
Specifically, we added offsets to the R17 abundances so that the abundance of CD$-48^{\circ}$02445 from our analysis and that from their analysis are consistent.
We note that the Na abundance of CD$-48^{\circ}$02445 was incorrectly reported in R17  and the correct value is [{Na}/{Fe}]$=-0.309$ (H. Reggiani 2021 private communication). 
We adopted this correct value to shift the R17 abundances. 
We also note that the metal-poor half of the R17 sample was analyzed relative to HD338529 and not to CD$-48^{\circ}$02445. 
Therefore, it is possible that R17 stars below [{Fe}/{H}]$\lesssim -2.1$ are not on the same abundance scale as the rest of the stars in R17 and stars in the present study and NS10.

The adopted abundances are summarized in Table~\ref{tab:abundance} for Sequoia stars and in Table~\ref{tab:abundance_standard} for the standard stars.
We note that the abundance of HD59392 listed in Table~\ref{tab:abundance_standard} was exactly the same as the one provided in NS10.

\subsection{Comparison with the literature}
\begin{figure*}
\centering
\includegraphics[width=1.0\textwidth]{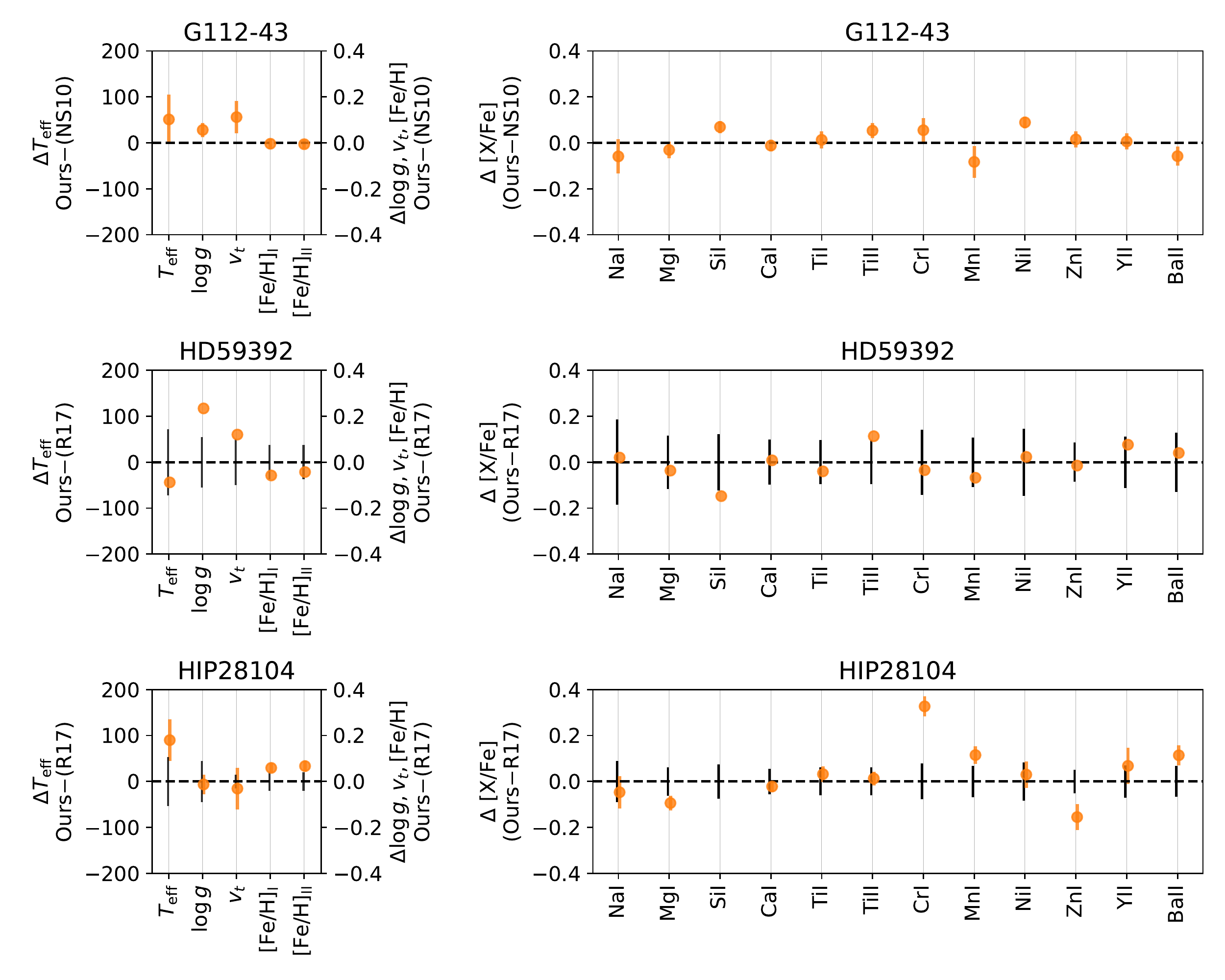}
\caption{Stellar parameter and abundance comparison with NS10 (G112-43) and R17 (HD59392 and HIP28104). The black error bars reflect the uncertainties reported in the literature. \label{fig:param_ab_comparison}}
\end{figure*}

\begin{figure}
\centering
\includegraphics[width=0.5\textwidth]{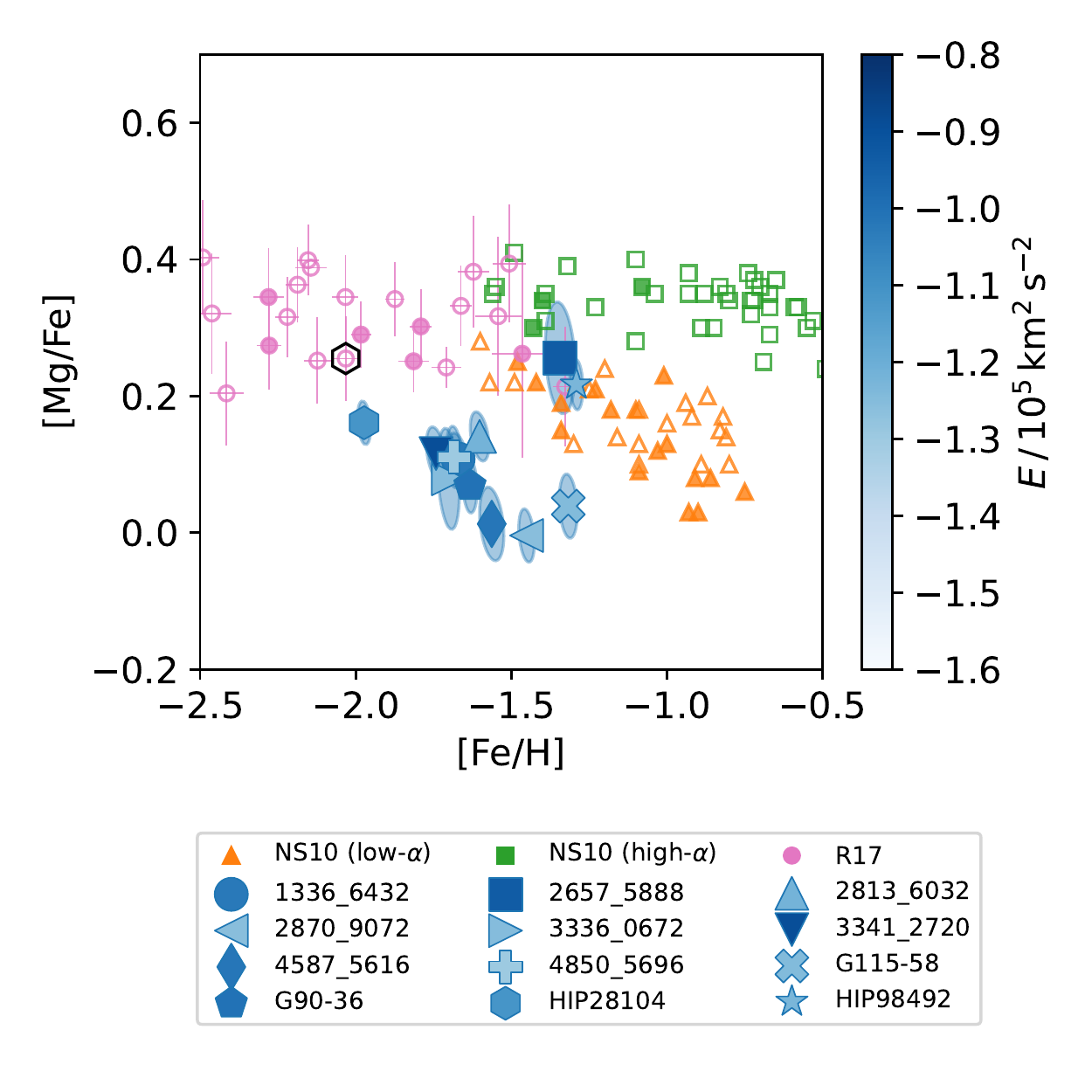}
\caption{Mg abundance of the stars. Kinematically selected Sequoia stars are plotted with blue symbols and color-coded according to their orbital energy (see Figure~\ref{fig:kinematicsCMD}). The measurement uncertainty and the covariance between [{Mg}/{Fe}] and [{Fe}/{H}] are indicated by the error ellipses. The data point circled with a black hexagon from R17 indicates the abundance of HIP28104. The abundance of the comparison samples comes from \citet{Nissen2010},\citet{Nissen2011}, and \citet{Reggiani2017a}. Filled symbols are for kinematically-selected Gaia-Enceladus stars among the comparison. \label{fig:Mg}}
\end{figure}

\begin{figure}
\centering
\includegraphics[width=0.4\textwidth]{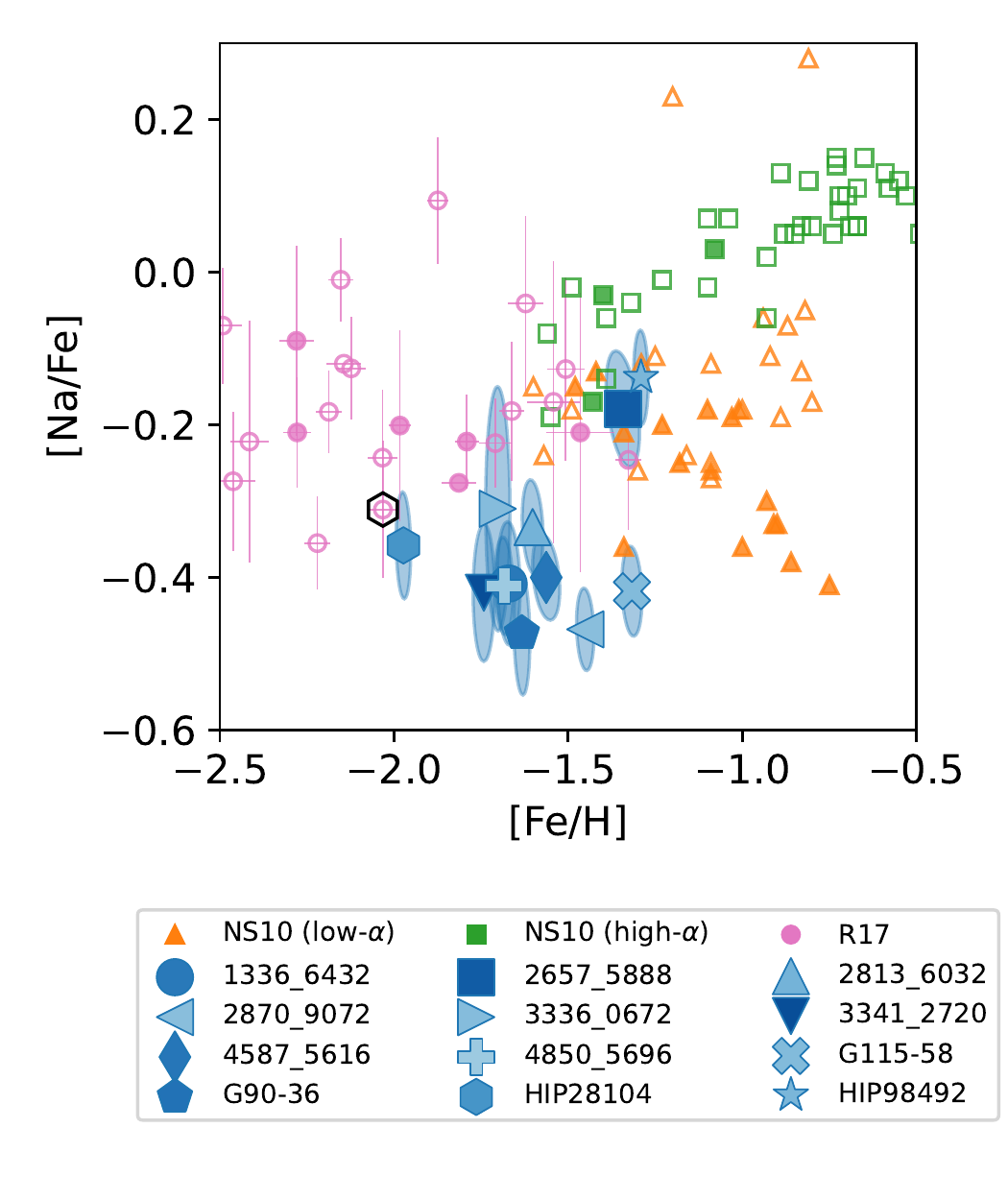}
\caption{Na abundances of the stars. Symbols are the same as those provided in Figure~\ref{fig:Mg}.\label{fig:Na}}
\end{figure}

\begin{figure*}
\centering
\includegraphics[width=1.0\textwidth]{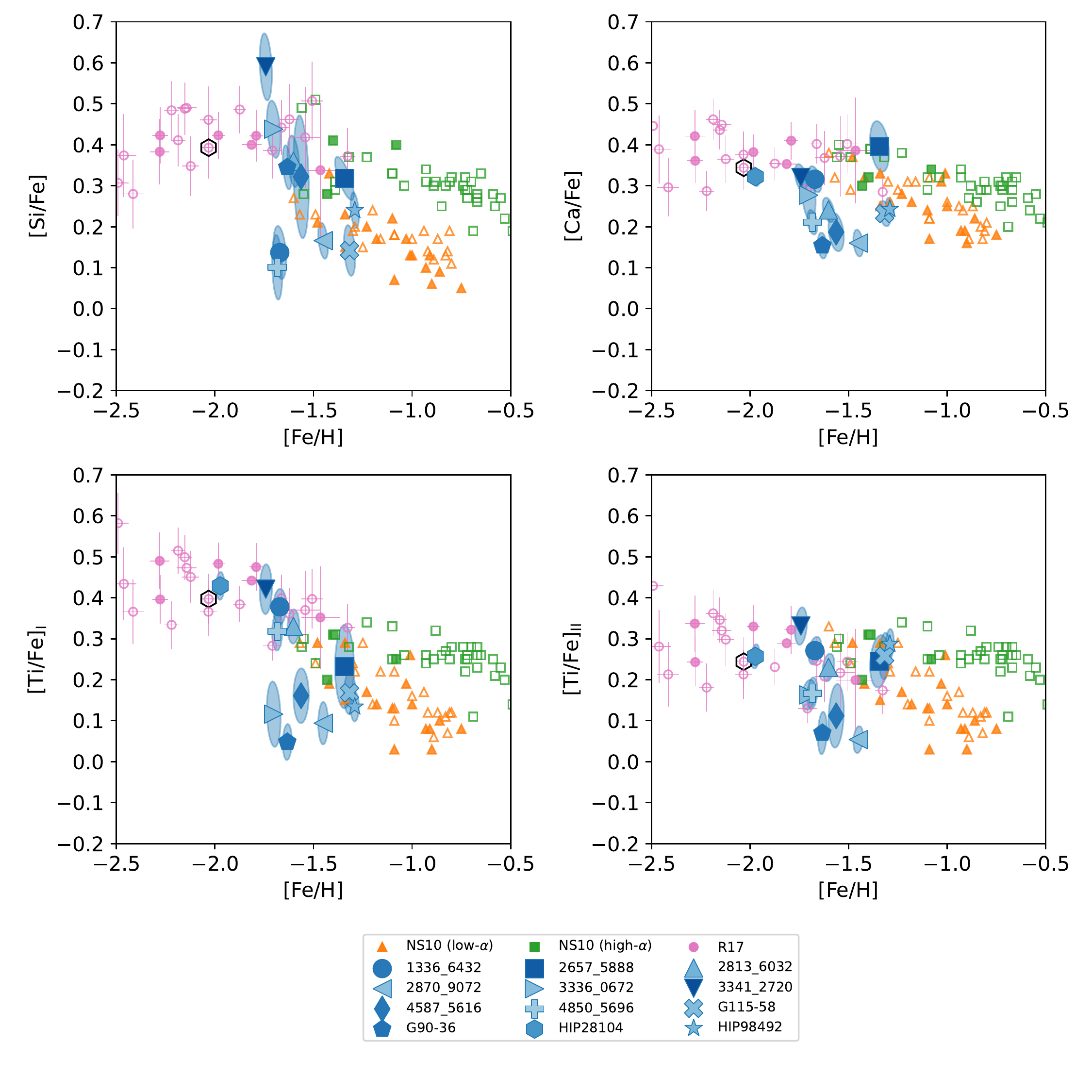}
\caption{Abundances of $\alpha$-elements of the stars. Symbols are the same as those provided
in Figure~\ref{fig:Mg}.\label{fig:alpha}}
\end{figure*}

In this section, we compare the stellar parameters and chemical abundances of standard stars with the literature.
For this purpose, we use G112-43 for the comparison with NS10 and HD59392 and HIP28104 for the comparison with R17.
This comparison is presented in Figure \ref{fig:param_ab_comparison}.

There is excellent agreement in the abundance of G112-43 with NS10. 
The difference in [{X}/{Fe}] is smaller than $0.1\,\mathrm{dex}$ in all of the elements.
The differences are smaller than two times our measurement uncertainties for most of the species.
We therefore consider that we successfully put our abundance onto NS10's scale using the standard star and that our uncertainty estimates are not underestimated.
The exceptions are Si and Ni, for which our abundances differ from NS10's measurements by more than a 2$\sigma$ uncertainty (2.3$\sigma$ and 3.6$\sigma$, respectively).
However, we note that we have not taken the measurement uncertainties in NS10 into account here since NS10 do not provide measurement uncertainties for individual objects.

We now compare our results with R17.
Since HD59392 is the standard star, there is no uncertainty in our abundance.
The difference between the adopted abundance of HD59392 is consistent with R17 within the uncertainties reported by R17.
Therefore, we have successfully put all the abundances onto the same scale for our study, NS10, and R17.
We note that HIP28104 shows large difference in [{X}/{Fe}], especially for Cr and Zn.
Since HIP28104 was not analyzed relative to CD$-48^{\circ}$02445 in R17 (see Section~\ref{sec:abundance}), the large offset is likely due to the use of different standard stars in R17.

\section{Results\label{sec:result}}

\begin{figure*}
\centering
\includegraphics[width=1.0\textwidth]{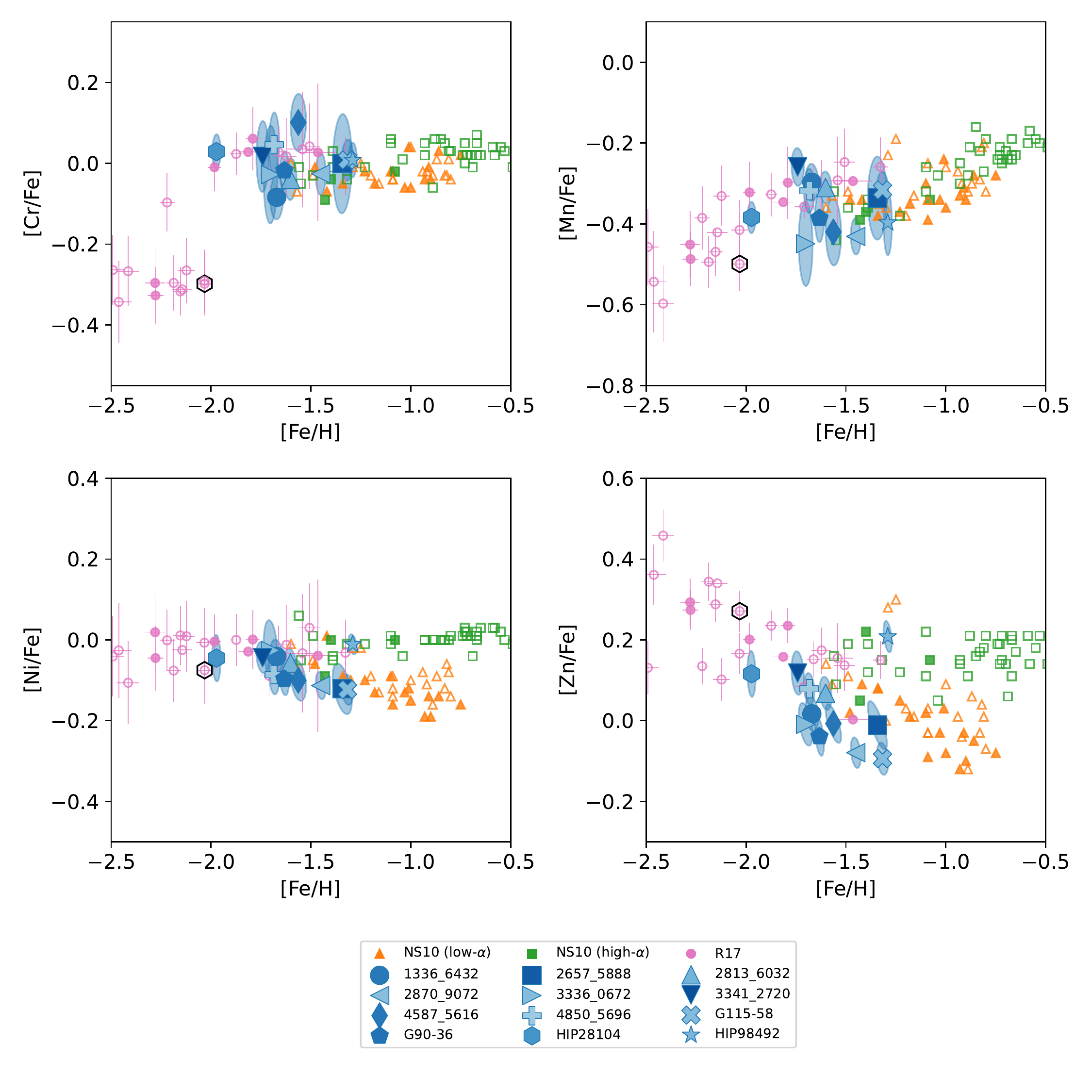}
\caption{Abundances of Cr, Mn, Ni, and Zn. Symbols are the same as those provided
in Figure~\ref{fig:Mg}.\label{fig:iron}}
\end{figure*}

\begin{figure*}
\centering
\includegraphics[width=0.8\textwidth]{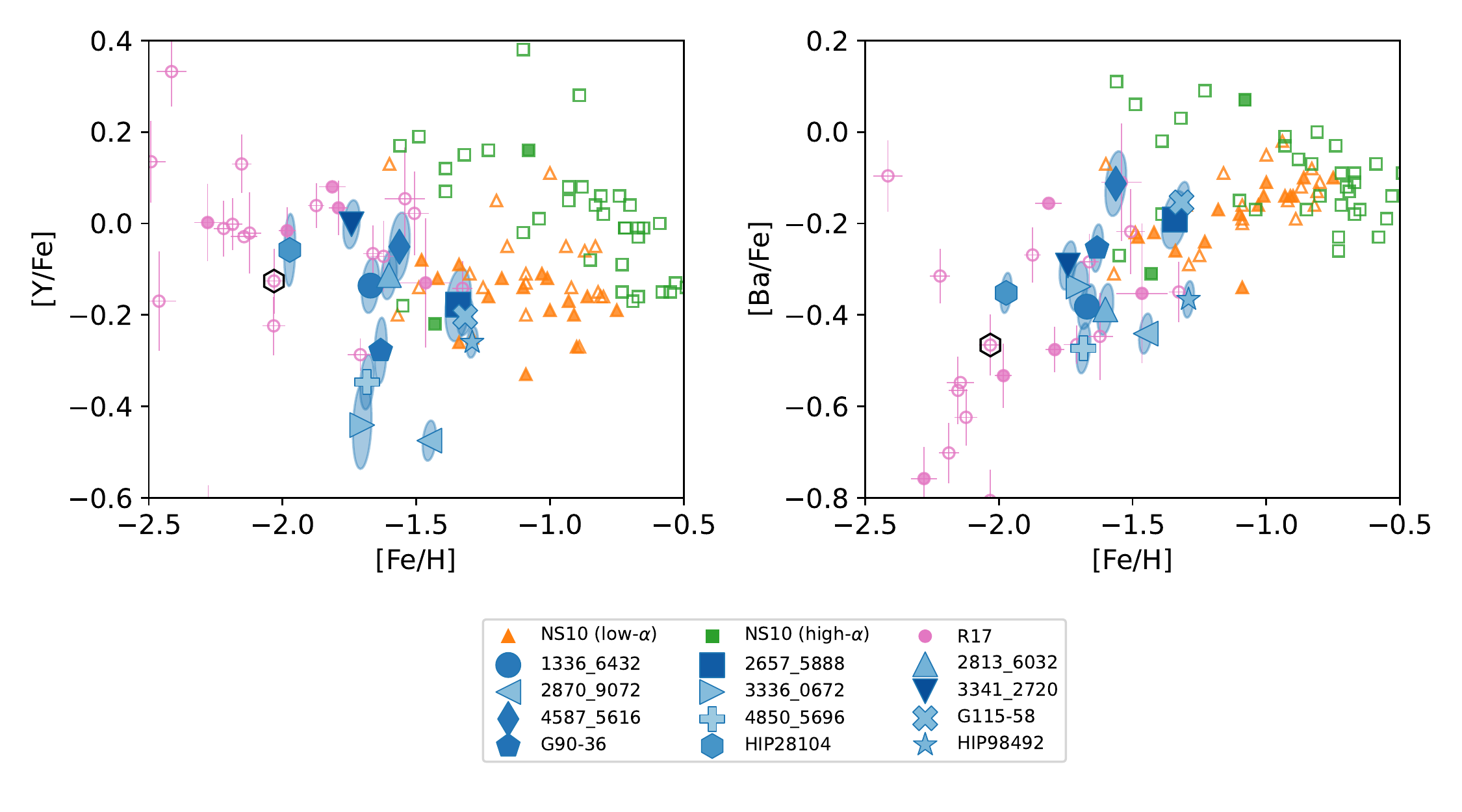}
\caption{Abundances of neutron-capture elements. Symbols are the same as those provided
in Figure~\ref{fig:Mg}.\label{fig:ncapture}}
\end{figure*}

In this section, we compare the chemical abundance of stars in the present study with those in NS10 and R17.
We kinematically selected Gaia-Enceladus stars out of NS10 and R17 samples using $L_z$ and $E_n$ as $|L_z|<600\,\mathrm{kpc\,km\,s^{-1}}$ and $-1.6<E/10^5\,\mathrm{km^2\,s^{-2}}<-1.0$ (Figure~\ref{fig:kinematicsCMD}).
We have 19 NS10 stars and six R17 stars that satisfy the Gaia-Enceladus kinematic selection. 
Among the 19 kinematically selected stars from NS10, 16 stars belong to their low-$\alpha$ population, confirming previous findings on the chemical abundance of Gaia-Enceladus. 
One of the remaining three stars is at $[\mathrm{Fe/H}]=-1.08$ and clearly has a high-$\alpha$ abundance, which is clearly different from the rest of the Gaia-Enceladus stars.
Thus, we removed this star from the Gaia-Enceladus sample.
We kept the remaining two stars from NS10's high-$\alpha$ population, since they are at $[\mathrm{Fe/H}]\sim-1.4$, where the low-$\alpha$ and high-$\alpha$ populations begin to overlap.
We thus ended up with 24 Gaia-Enceladus stars from the literature, which provide us with opportunities to compare abundance patterns of our Sequoia stars with those of Gaia-Enceladus stars.

The other stars in NS10 and R17 allow us to compare Sequoia stars with in situ (hot thick disk) stars and other typical halo stars. 
The high-$\alpha$ population in NS10 are considered to be stars that formed in situ; they are likely to have formed in the Milky Way and to have later heated onto the halo-like orbits. 
We simply call these stars ``in situ stars'' in what follows.
The origin of R17 stars and NS10's low-$\alpha$ stars that do not satisfy the Gaia-Enceladus selection is less clear. 
They are either Gaia-Enceladus stars, in situ stars, or stars from accreted galaxies other than Gaia-Enceladus.
As there are no strong kinematic selections in NS10 and in R17, we consider these stars as typical halo stars.

We here compare chemical abundances of kinematically selected Sequoia stars, NS10 and R17 stars.
We start the comparison with [{Mg}/{Fe}] in Figure~\ref{fig:Mg} since NS10 used this abundance ratio to separate high- and low-$\alpha$ populations.
The majority of Sequoia stars do not follow the sequence formed by Gaia-Enceladus and they have lower [{Mg}/{Fe}] ratios than Gaia-Enceladus stars at the same metallicity.
A similar difference between Sequoia and Gaia-Enceladus stars is also found for Na (Figure~\ref{fig:Na}) and some $\alpha$ elements (Figure~\ref{fig:alpha}).
It is hard to see the difference in elemental abundances near the iron peak, except for Zn (Figure~\ref{fig:iron}).
One of the neutron capture elements, Y, also shows a different behavior between Sequoia and Gaia-Enceladus stars (Figure~\ref{fig:ncapture}).

In summary, although a few objects that are kinematically selected as part of Sequoia at high metallicity (2657\_5888 and HIP98492) seem to follow the trend of Gaia-Enceladus, there are clearly differences between the majority of Sequoia stars and Gaia-Enceladus stars in [{Mg}/{Fe}], [{Na}/{Fe}], [{Ca}/{Fe}], [{Ti}/{Fe}], [{Zn}/{Fe}], and [{Y}/{Fe}].
Our homogeneously derived high-precision chemical abundances robustly confirm the finding of \citet{Matsuno2019a} in Na, Mg, and Ca; \citet{Monty2020a} in Mg and Ca; and \citet{Aguado2021a} in the overall $\alpha$-element abundance.   

\begin{figure}
\centering
\includegraphics[width=0.4\textwidth]{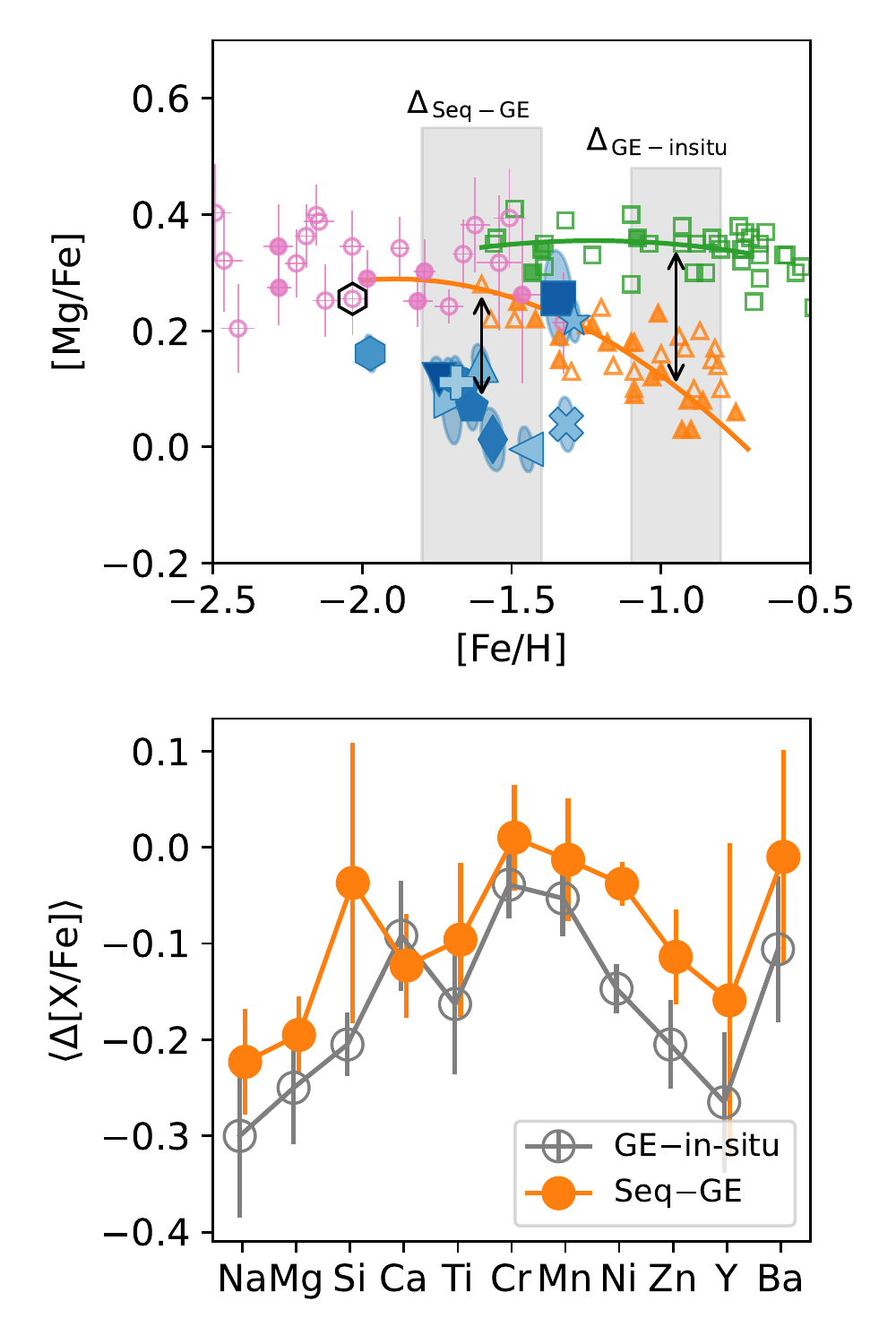}
\caption{Comparison of patters in abundance difference. (Top:) Visualization of the calculations of $\Delta_{\rm Seq-GE}$ and $\Delta_{\rm GE-in-situ}$. See text for details. (Bottom:) Mean abundance difference between Sequoia and Gaia-Enceladus at $-1.8<[\mathrm{Fe/H}]<-1.4$, and that between Gaia-Enceladus and in situ stars at $-1.1<[\mathrm{Fe/H}]<-0.8$ ($\langle \Delta_{\rm Seq-GE}\rangle$ and $\langle \Delta_{\rm GE-in-situ}\rangle$). The error bar represents the standard deviation. \label{fig:GEseq}}
\end{figure}

In order to quantify these differences, we followed the approach of \citet{Nissen2011} (see top panel of Figure \ref{fig:GEseq}).
We first fit the abundance trend of Gaia-Enceladus in [{X}/{Fe}]--[{Fe}/{H}] with a quadratic polynomial using the kinematically selected stars and calculate the residual scatter ($\sigma_{\rm resid}$). 
For each of the eight Sequoia stars in the metallicity range $-1.8<[\mathrm{Fe/H}]<-1.4$, we computed the deviation from this fit ($\Delta_{\mathrm{Seq-GE}}$). 
We then computed $\chi^2=\sum \Delta^2_{\mathrm{Seq-GE}}  / (\sigma_{\rm resid}^2+\sigma^2([\mathrm{X/Fe}]))$ and conducted a $\chi^2$ test to obtain the probability that a $\chi^2$ distribution with the degree of freedom of eight has a $\chi^2$ higher than the observed value.
This is to test if we can explain the displacement in abundance ratios of Sequoia stars from the trend of Gaia-Enceladus stars with the residual in the fit and the measurement uncertainties. 

We also computed the abundance difference between Gaia-Enceladus and in situ stars ($\Delta_{\mathrm{GE-in{\text -}situ}}$) at $-1.1<[\mathrm{Fe/H}]<-0.8$ by fitting the abundance trend of in situ stars in the same way. 
This calculation basically provides a similar quantity as Table 5 of \citet{Nissen2011}.
A difference to \citet{Nissen2011} is that here we compare kinematically selected Gaia-Enceladus stars instead of the low-$\alpha$ population with in situ stars (high-$\alpha$ population). 

The results are summarized in Table~\ref{tab:chitest}.
The average abundance difference between Gaia-Enceladus and Sequoia is largest in [{Na}/{Fe}] (Figure~\ref{fig:Na}), which is followed by [{Mg}/{Fe}] (Figure \ref{fig:Mg}).  
The differences in these abundance ratios are $\sim 0.2\,\mathrm{dex}$ and highly significant.
Other $\alpha$ elements, Ca and Ti, show differences of $\sim 0.1\,\mathrm{dex}$ in [{X}/{Fe}], while the difference in [{Si}/{Fe}] is not as large as the other $\alpha$ elements (see Figure~\ref{fig:alpha}). 
Although the results of the $\chi^2$ tests are significant for all of Si, Ca, and Ti, the large $\chi^2$ in [{Si}/{Fe}] might not be due to the average abundance difference between Sequoia and Gaia-Enceladus, but to the large spread in [{Si}/{Fe}] ratios in our Sequoia stars. 
Given that we had to rely on one or a few weak Si lines for its abundance determination, the [{Si}/{Fe}] distribution needs to be taken with a caution.

These $\alpha$-element abundance differences between Sequoia and Gaia-Enceladus are quite similar to those found by \citet{Nissen2011} between their low-$\alpha$ and high-$\alpha$ populations, which are still seen when we compare Gaia-Enceladus and the high-$\alpha$ in situ population (Table~\ref{tab:chitest}).
Even though the differences are found at a different metallicity, this indicates that the same physical mechanism, specifically type~Ia supernovae
(SNe~Ia) in this case, is  likely responsible for creating the abundance differences. 
We discuss this further in Section~\ref{sec:discussion}.

Iron-group elements including Cr, Mn, and Ni do not show a significant abundance difference between Sequoia and Gaia-Enceladus (see Figure~\ref{fig:iron}), although there is clearly a difference in [{Ni}/{Fe}] between Gaia-Enceladus and in situ stars \citep[Table~\ref{tab:chitest};][]{Nissen2011}.
The Ni abundance difference between Sequoia and Gaia-Enceladus is $\Delta[\mathrm{Ni/Fe}]\sim 0.04\,\mathrm{dex}$ if present.
On the other hand, there is a statistically significant difference in [{Zn}/{Fe}] by $\sim 0.11\,\mathrm{dex}$ between Sequoia and Gaia-Enceladus (Table~\ref{tab:chitest}).
Zinc also shows abundance differences between Gaia-Enceladus and in situ stars.

Neutron capture elements (Y and Ba) show relatively large scatters compared to other elements (Figure~\ref{fig:ncapture}, see also Table~\ref{tab:chitest}). 
Despite the large scatter, [{Y}/{Fe}] is significantly smaller for  Sequoia stars than for Gaia-Enceladus stars (Table~\ref{tab:chitest}).
Furthermore, Y is another element that shows an abundance difference in [{X}/{Fe}] between Gaia-Enceladus and in situ stars.

\begin{table*}
\caption{Abundance difference between Sequoia and Gaia-Enceladus at $-1.8<[\mathrm{Fe/H}]<-1.4$, and that between Gaia-Enceladus and in situ stars at $-1.1<[\mathrm{Fe/H}]<-0.8$. \label{tab:chitest}}
\centering
\begin{tabular}{l*{6}{r}}
\hline\hline
Species & \multicolumn{2}{c}{$\Delta_{{\rm Sequoia}-{\rm GE}}$} & $\sigma([\mathrm{X/Fe}])$ &\multicolumn{2}{c}{$\Delta_{{\rm GE} -{\rm in-situ}}$} & $p_{\chi^2}$ \\\cline{2-3}\cline{5-6}
        &          Mean  &          Std. & Median  &  Mean        &      Std.&           \\\hline

NaI&-0.223&0.055&0.073&-0.300&0.085&0.000\\
MgI&-0.195&0.040&0.040&-0.250&0.059&0.000\\
SiI&-0.037&0.146&0.068&-0.205&0.033&0.005\\
CaI&-0.123&0.054&0.033&-0.092&0.057&0.000\\
TiI&-0.113&0.119&0.050&-0.162&0.073&0.000\\
TiII&-0.096&0.080&0.038&-0.163&0.073&0.005\\
CrI&0.010&0.055&0.063&-0.039&0.035&0.843\\
MnI&-0.013&0.064&0.051&-0.053&0.040&0.552\\
NiI&-0.038&0.023&0.041&-0.147&0.026&0.725\\
ZnI&-0.114&0.049&0.045&-0.205&0.046&0.004\\
YII&-0.159&0.163&0.060&-0.265&0.073&0.000\\
BaII&-0.010&0.111&0.053&-0.106&0.076&0.258\\\hline

\end{tabular}
\end{table*}

We finally mention the relation between kinematics and chemical abundances within Sequoia. 
In Figures~\ref{fig:Mg}--\ref{fig:ncapture}, the orbital energy of the stars are indicated by the intensity of the color.
We did not find any significant correlations between kinematics and chemical abundances among Sequoia stars. 
We conducted $t$ tests for the hypothesis that Sequoia stars, having high energy ($E>1.2\times10^5\,\mathrm{km^2\,s^{-2}}$), have the same [{X}/{Fe}] as lower energy Sequoia stars at the metallicity range of $-1.8<[\mathrm{Fe/H}]<-1.4$.
In all species, we cannot reject the hypothesis at more than a $2\sigma$ level, indicating that there are no significant abundance differences between Sequoia stars at a high energy and at a low energy.

\section{Discussion\label{sec:discussion}}
\subsection{Chemical enrichments in Sequoia}

\begin{figure*}
\centering
\includegraphics[width=1.0\textwidth]{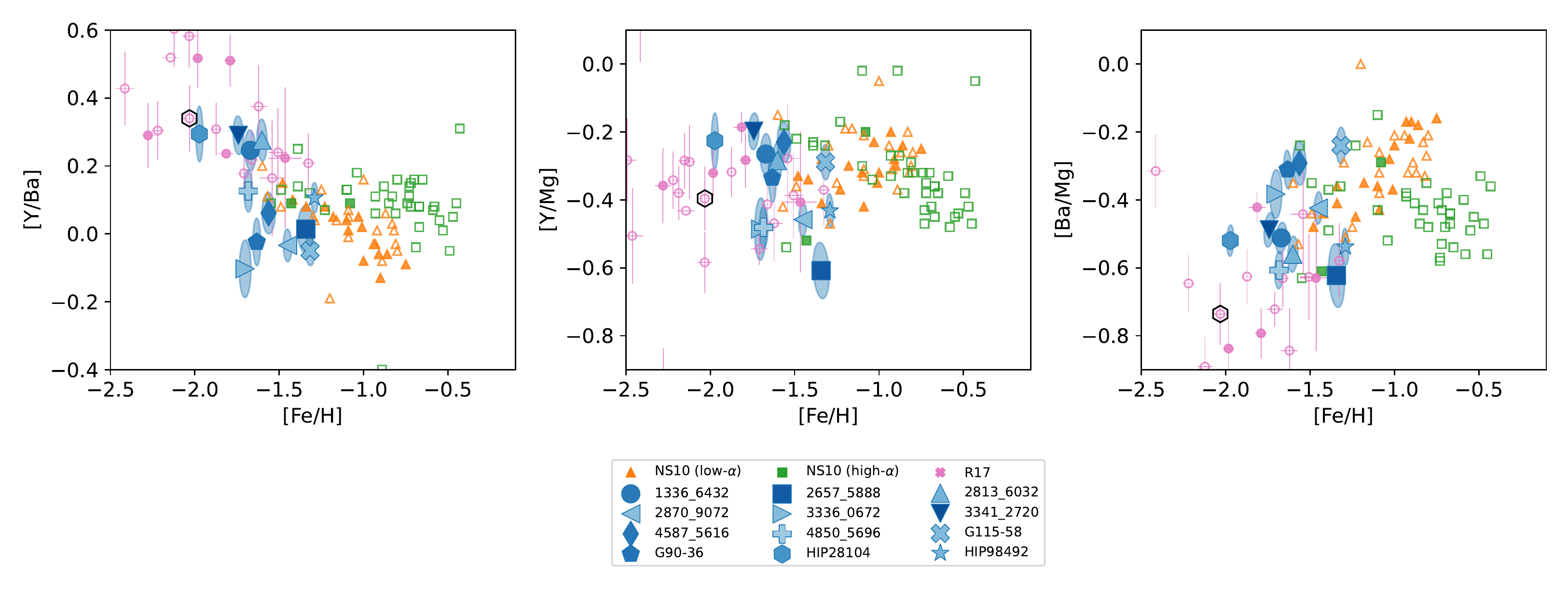}
\caption{Neutron-capture element abundances. Symbols are the same as those provided in Figure~\ref{fig:Mg}. \label{fig:BaY}}
\end{figure*}

The majority of the kinematically selected Sequoia stars that we have studied seem to show a different chemical abundance than Gaia-Enceladus, indicating that their progenitor is indeed different from Gaia-Enceladus. 
There are also a few stars that seem to have a chemical abundance comparable to Gaia-Enceladus. 
Although they could be stars stripped in the early phase of the Gaia-Enceladus accretion, as suggested by \citet{Koppelman2019a}, they do not seem to be a dominant population among stars having large retrograde motion and small binding energy.
 In the next section, we discuss which stars are likely true members of Sequoia from the point of chemical abundance.

In this section, we focus on the majority of stars having a different abundance pattern than Gaia-Enceladus and we further discuss the origin of the chemical signature of Sequoia.
Sequoia stars tend to show lower [{X}/{Fe}] in the same elements that show deficiency in Gaia-Enceladus in comparison to the in situ stars at high metallicity.
We visualize this similarity in Figure~\ref{fig:GEseq}, where the mean abundance difference between Sequoia and Gaia-Enceladus stars at $-1.8<[\mathrm{Fe/H}]<-1.4$ ($\langle \Delta_{\rm Seq-GE} \rangle$) from Table~\ref{tab:chitest} is plotted. 
The error bar reflects the standard deviation in [{X}/{Fe}] in Sequoia. 
We also plotted the mean abundance difference between Gaia-Enceladus and in situ stars at $-1.1<[\mathrm{Fe/H}]<-0.8$ ($\langle \Delta_{\rm GE-in-situ} \rangle$).

Figure~\ref{fig:GEseq} demonstrates a remarkable similarity between the patterns in $\langle\Delta_{\rm Seq-GE}\rangle$ and $\langle\Delta_{\rm GE-in-situ}\rangle$. 
We find the largest differences for Na and Mg, and mild differences for Ca, Ti, Zn, and Y in both comparisons.
Although $\langle\Delta_{\rm GE-in-situ}\rangle$ is clearly nonzero and negative in Si and Ni, it is hard to see a similar feature in $\langle\Delta_{\rm Seq-GE}\rangle$ for these elements.  
The lack of difference in the Si abundance comparison between Sequoia and Gaia-Enceladus could be due to our larger measurement uncertainty and smaller number of stars compared to NS10 and \citet{Nissen2011}.
In addition, Cr and Mn do not show significant differences in both comparisons.

The similarity in $\langle \Delta_{\rm Seq-GE} \rangle$ and $\langle \Delta_{\rm GE-in-situ}\rangle$ could suggest that the same physical process is responsible for shaping their patterns.
The origin of the abundance difference between Gaia-Enceladus and in situ stars is usually attributed to chemical enrichment by SNe~Ia;
while in situ stars have not been significantly enriched by SNe~Ia at [{Fe}/{H}]$\sim -1$ yet, Gaia-Enceladus has already been enriched by SNe~Ia because of a longer star formation timescale, a lower star formation efficiency, and/or different star formation histories \citep[e.g.,][]{Zolotov2010,Nissen2010,Nissen2011,FernandezAlvar2018a,Gallart2019a,Brook2020a,Sanders2021a}, which would be a consequence of the lower halo mass of Gaia-Enceladus than the main progenitor of the Milky Way.
If we apply the same reasoning, the chemical abundance difference between Sequoia and Gaia-Enceladus is likely caused by larger SNe~Ia enrichments in Sequoia at [{Fe}/{H}]$\sim -1.5$, which would indicate a lower mass for Sequoia than Gaia-Enceladus.
This is consistent with the mass estimates from kinematic analysis, metallicity distribution, or a number count of stars \citep{Koppelman2019a,Myeong2019a,Matsuno2019a,Naidu2020a}.

Although $\langle \Delta_{\rm Seq-GE} \rangle$ and $\langle \Delta_{\rm GE-in-situ}\rangle $ show a remarkable overall similarity, one may notice a slight difference in Figure~\ref{fig:GEseq}, especially in Ni.
If the similarity in patterns in $\langle \Delta_{\rm Seq-GE} \rangle$ and $\langle \Delta_{\rm GE-in-situ}\rangle $ is primarily driven by the chemical enrichment from SNe~Ia, the lack of a Ni abundance difference between Sequoia and Gaia-Enceladus might indicate a variation in the properties of SNe~Ia with, for example, environments or metallicity.
For instance, \citet{Kirby2019a} and \citet{Sanders2021a} used a Ni abundance to conclude that the explosion of sub-Chandrasekhar mass is the dominant type of SNe~Ia in dwarf galaxies and in Gaia-Enceladus.
The evolution of a Ni abundance is actually sensitive to the explosion mechanism of a dominant type of SNe~Ia \citep{Palla2021a}.

We note that, in addition to the larger contribution of SNe~Ia, \citet{FernandezAlvar2019a} suggest a possibility of a top light initial mass function for Gaia-Enceladus by modeling $\alpha$-element abundances of halo stars from APOGEE.
It remains to be seen if a similar explanation can be applied to Sequoia.  

We now focus on neutron-capture elements, Y and Ba.
While both elements are produced by the $s$ process at solar metallicity, Y belongs  to a group of light $s$-process elements.
The weak $s$ process, which operates in massive stars, contributes more to light $s$-process elements than the main $s$ process in low-to-intermediate mass stars.
In addition, the $r$ process also significantly contributes to the enrichments of these elements in the early universe.
Yttrium shows a relatively large scatter and a mild deficiency in Sequoia (Table~\ref{tab:chitest}). 
The Y deficiency seems to be driven by a few stars with low [{Y}/{Fe}].
They have [{Y}/{Fe}]$\sim -0.4$ and such a low abundance is hardly seen among Gaia-Enceladus stars in the NS10 and R17 samples.
Barium also shows a large scatter, although the average abundance seems to be comparable to Gaia-Enceladus stars in the comparison samples.

To understand these trends, we plotted [{Y}/{Ba}], [{Y}/{Mg}], and [{Ba}/{Mg}] abundances in Figure~\ref{fig:BaY}.
The [{Y}/{Ba}] ratio allowed us to infer the importance of a weak $s$ process relative to the efficiency of Ba production either by low-to-intermediate mass stars or by $r$-process nucleosynthesis.
The [{Y}/{Ba}] ratios of Sequoia stars tend to be lower than those of Gaia-Enceladus stars. 
The average difference between the two systems is $-0.15\,\mathrm{dex}$.
Gaia-Enceladus is also known to possess low [{Y}/{Ba}] values compared to in situ stars at [{Fe}/{H}]$\sim -1$ \citep{Nissen2011}. 

The low [{Y}/{Ba}] can be understood if the efficiency of the weak $s$ process is lower in Sequoia, if low-to-intermediate mass stars in which the main-$s$ process operates start to contribute to the chemical evolution of Sequoia, or if there are more $r$-process nucleosynthesis events in Sequoia that produce more Ba than Y.
The abundance ratios [{Y}/{Mg}] and [{Ba}/{Mg}] displayed in the right two panels of Figure~\ref{fig:BaY} further allowed us to infer the efficiency of the enrichment of the element X relative to the chemical enrichment by core-collapse supernovae (CCSNe) since CCSNe produce most of the Mg. 
Sequoia stars stand out less in [{Y}/{Mg}], indicating that the production of Y is controlled by the abundance of Mg to some extent. 
This is expected if the majority of Y is produced by the weak $s$ process, since its efficiency is dependent on CNO abundances \citep[e.g.,][]{Prantzos1990a}, which are mostly produced by massive stars.
A similar argument is applied to explain the Cu (and Y) abundance differences between Gaia-Enceladus and in situ stars \citep{Nissen2011,Matsuno2021a}.
Sequoia stars, on the other hand, are slightly enhanced in [{Ba}/{Mg}], which might be related to the high Eu abundance reported by \citet{Aguado2021a} and hence indicate efficient $r$-process nucleosynthesis in Sequoia.

While Y and Ba abundances provide some insights about the enrichment of neutron-capture elements in Sequoia, the information is still limited.
It is highly desirable to measure abundances of many neutron-capture elements in future studies.
For example, \citet{Aguado2021a} suggest a possibility of enhanced $r$-process element abundance in Gaia-Enceladus and Sequoia from an Eu abundance, and \citet{Matsuno2021b} explain the high Eu abundance of Gaia-Enceladus as a combined effect of delay time in the $r$-process enrichment and the low star formation efficiency of Gaia-Enceladus.
In a forthcoming paper, we plan to revisit weak and main $s$ processes, and $r$-process enrichments in Sequoia with precise abundances of more neutron-capture elements (e.g., Sr, Eu).

There are some similarities in abundance ratios between Sequoia and the surviving dwarf galaxies around the Milky Way, such as Sagittarius, Fornax, Draco, Sculptor, Sextans dwarf spheroidal galaxies, and the Large and Small Magellanic Clouds (LMC and SMC).
The Milky Way, Gaia-Enceladus, and many of the surviving dwarf galaxies show similarly super-solar [{$\alpha$}/{Fe}] at low metallicity. 
However, since the ''knee'' metallicity ($[\mathrm{Fe/H}]_{\rm knee}$) at which systems start to show decreasing [{$\alpha$}/{Fe}] with metallicity is below $[\mathrm{Fe/H}]\lesssim -1.8$ in Fornax, Draco, Sculptor, and Sextans dwarf galaxies, they show lower [{$\alpha$}/{Fe}] than Gaia-Enceladus or the Milky Way in situ stars at $[\mathrm{Fe/H}]_{\rm knee}<[\mathrm{Fe/H}]$ \citep{Tolstoy2009,Cohen2009a,Kirby2011a,Lemasle2012,Lemasle2014,Hendricks2014a,Hill2019a,Theler2020a}.
Although the position of the knee is less clear in the LMC, SMC, and Sagittarius, they also show low $\alpha$-element abundance at $[\mathrm{Fe/H}]\sim -1.5$ \citep{Nidever2020a,Hasselquist2021a}.
It is not clear if Sequoia shows a clear knee because of insufficient sampling of low-metallicity stars.
Nonetheless, the abundance of the most metal-poor star in our sample (HIP28104) seems to support the existence of the knee at low metallicity.

\subsection{Chemical identification of Sequoia members\label{sec:chemicalseparation}}

\begin{figure}
\centering
\includegraphics[width=0.5\textwidth]{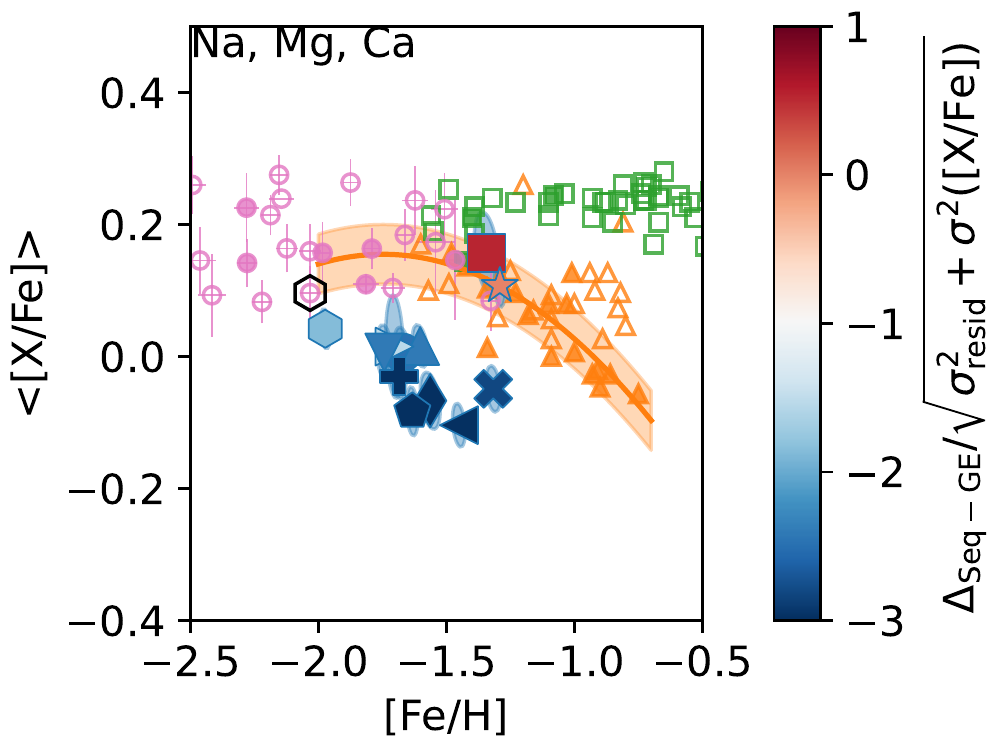}
\caption{Average values of $[\mathrm{X/Fe}]$ in Na, Mg, and Ca are plotted against [{Fe}/{H}]. For the calculation of the uncertainty in $\langle[\mathrm{X/Fe}]\rangle$, we properly take correlated uncertainties between elemental abundances into account. Stars are color-coded according to the significance of their departure from the Gaia-Enceladus sequence, which is shown as the orange sold line. The Gaia-Enceladus sequence was obtained by fitting the abundance ratios of Gaia-Enceladus stars from \citet{Nissen2010} and \citet{Reggiani2017a} with a quadratic polynomial. The residual scatter of Gaia-Enceladus around the fit is also shown around the fit.
\label{fig:meanXFE}}
\end{figure}

Under the assumption that each system shows a well-defined track in [{X}/{Fe}]--[{Fe}/{H}] planes and there are no chemical outliers, we should be able to chemically identify members of disrupted galaxies based on chemical abundance ratios.
In this section, we try to separate individual Sequoia stars from Gaia-Enceladus ones solely based on the abundance ratios. 

We first need to explore what is the best combination of elements to identify Sequoia stars.
While considering the average of more elemental abundance ratios might lead to a reduced uncertainty, abundance differences might be smeared out if we add elements whose abundance does not show a difference between Sequoia and Gaia-Enceladus.
We considered an average abundance of various combinations out of five elements (Na, Mg, Ca, Ti from Ti~II, and Zn) since they show a clear abundance difference between Sequoia and Gaia-Enceladus in our study (Table~\ref{tab:chitest}).
To quantitatively identify the best combination of elements, we followed the approach that is explained in Figure~\ref{fig:GEseq}; we first fit the abundance trend of Gaia-Enceladus and then calculated $\chi^2$ values for Sequoia stars in $-1.8<[\mathrm{Fe/H}]<-1.4$.

The $\chi^2$ value was at its maximum when we considered the three elements, Na, Mg, and Ca (Figure~\ref{fig:meanXFE}).
Therefore, we consider that the average of abundance ratios of [{Na}/{Fe}], [{Mg}/{Fe}], and [{Ca}/{Fe}] offers a powerful diagnostic to chemically identify Sequoia stars.
We note that the best combination of abundance ratios, however, varies depending on the data set, specifically the typical uncertainties in abundances of elements. 

Among our sample, eight stars (1336\_6432, 2813\_6032, 2870\_9072, 3341\_2720, 4587\_5616, 4850\_5696, G115-58, and G90-36) deviate more than $2\sigma$ from the Gaia-Enceladus sequence in $\langle [\mathrm{Na/Fe}],[\mathrm{Mg/Fe}],[\mathrm{Ca/Fe}]\rangle$, where $\sigma$ is defined as $\Delta_{\rm Seq-GE}/\sqrt{\sigma_{\rm resid}^2+\sigma^2(\langle[\mathrm{X/Fe}]\rangle)}$. 
From the chemical point of view, these are the most likely members of Sequoia. 

On the other hand, four stars do not deviate by more than $2\sigma$.
Two stars (2657\_5888 and HIP98492) are at high metallicity and seem to be on the Gaia-Enceladus sequence. 
These stars might be those stripped in the very early stage of the Gaia-Enceladus accretion \citep{Koppelman2019a}.
The other two stars, 3336\_0572 and HIP28104, have a relatively low metallicity. 
Since 3336\_0572 is one of the stars for which the uncertainties in elemental abundances are large, the lack of significance might just reflect insufficient precision.
HIP28104 has the lowest metallicity among our sample. 
Its metallicity might be too low even for the progenitor of Sequoia to be affected by SNe~Ia. 

Using these numbers, we could constrain the contribution of Sequoia stars and Gaia-Enceladus stars stripped early \citep{Koppelman2019a} among those selected to be highly retrograde.
We found that two to four out of the 12 stars are from Gaia-Enceladus.
Taking the Poisson error into account, we estimated that Gaia-Enceladus stars can contribute to 5-50 \% of stars at large negative $L_z$ and high $E_n$.

We now assess the minimum precision ($\sigma_{\rm obs}$) required to separate individual Sequoia stars from Gaia-Enceladus using [{Mg}/{Fe}] or $\langle [\mathrm{Na/Fe}],[\mathrm{Mg/Fe}],[\mathrm{Ca/Fe}]\rangle$.
The scatter among Sequoia stars (Table~\ref{tab:chitest}) and the residual scatter of Gaia-Enceladus stars from the fitting are comparable to the measurement uncertainty ($\sim0.04\,\mathrm{dex}$ in both cases).
Therefore, we consider that the dispersion in abundance ratios among Sequoia stars would be dominated by the measurement uncertainty, $\sigma_{\rm obs}$.
The difference between Sequoia and Gaia-Enceladus in these abundance ratios ($|\Delta|$) are $\sim 0.20\,\mathrm{dex}$ and $\sim 0.18\,\mathrm{dex}$, respectively.
The required precision to chemically separate 84\% of Sequoia stars with the $2\sigma$ criterion can be calculated by solving 
\begin{equation}
|\Delta| - \sigma_{\rm obs}>2\sqrt{\sigma_{\rm resid}^2+\sigma_{\rm obs}^2},
\end{equation}
where the left side of the equation reflects the fact that about 84\% of Sequoia stars have a larger deviation in abundance ratios from the Gaia-Enceladus trend than this value.
Assuming $|\Delta|=0.19$ and $\sigma_{\rm resid}=0.04$, we obtained $\sigma_{\rm obs}\lesssim 0.07\,\mathrm{dex}$ from this equation.
This condition is met in our case even if we were to work only on Mg since the typical precision is $\sigma([\mathrm{Mg/Fe}])\sim 0.04\,\mathrm{dex}$. 
Considering more elemental abundance might help in some cases, although the typical uncertainty does not significantly improve in our case when we considered the three elements (Na, Mg, and Ca).
Here, we note that it is also important to consider correlations between abundances of the elements to correctly estimate the uncertainty for their average abundance.
 
On the other hand, it is much easier to detect the average difference in chemical abundance between Sequoia and Gaia-Enceladus stars.
The uncertainty of the mean is scaled with $\sqrt{N}$ when systematic uncertainty can be neglected.
Therefore, even if the observational uncertainty is comparable to the abundance difference between Sequoia and Gaia-Enceladus, four stars would be sufficient to detect the average abundance difference if the sample does not contain chemical outliers or contaminants.

\subsection{Sequoia stars in the literature and surveys}

\begin{figure}
\centering
\includegraphics[width=0.5\textwidth]{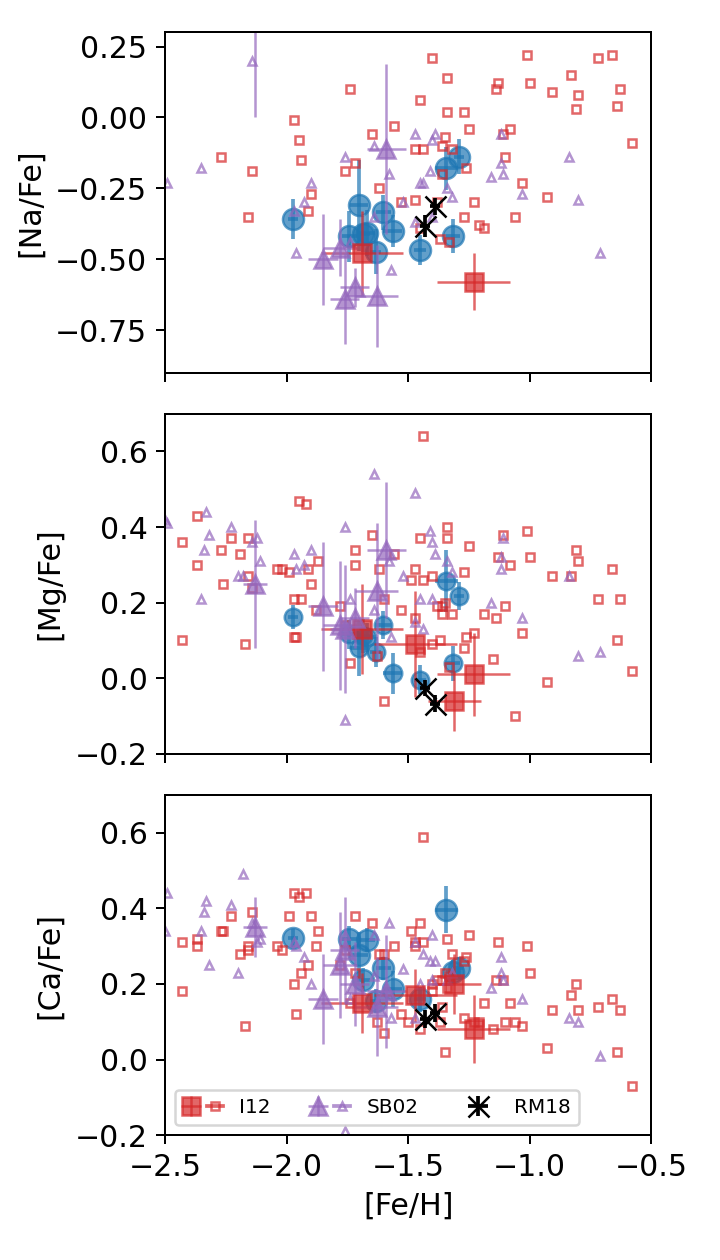}
\caption{Abundance of stars from the literature and from this study. The literature data come from \citet[][SB02]{Stephens2002}, \citet[][I12]{Ishigaki2012}, and \citet[][RM18]{Reggiani2018a}. Sequoia stars are shown with filled symbols. Stars in the present study are shown with blue circles. We note that the uncertainty is shown with error bars instead of error ellipses for visualization. \label{fig:seq_lite}}
\end{figure}

\begin{figure}
\centering
\includegraphics[width=0.5\textwidth]{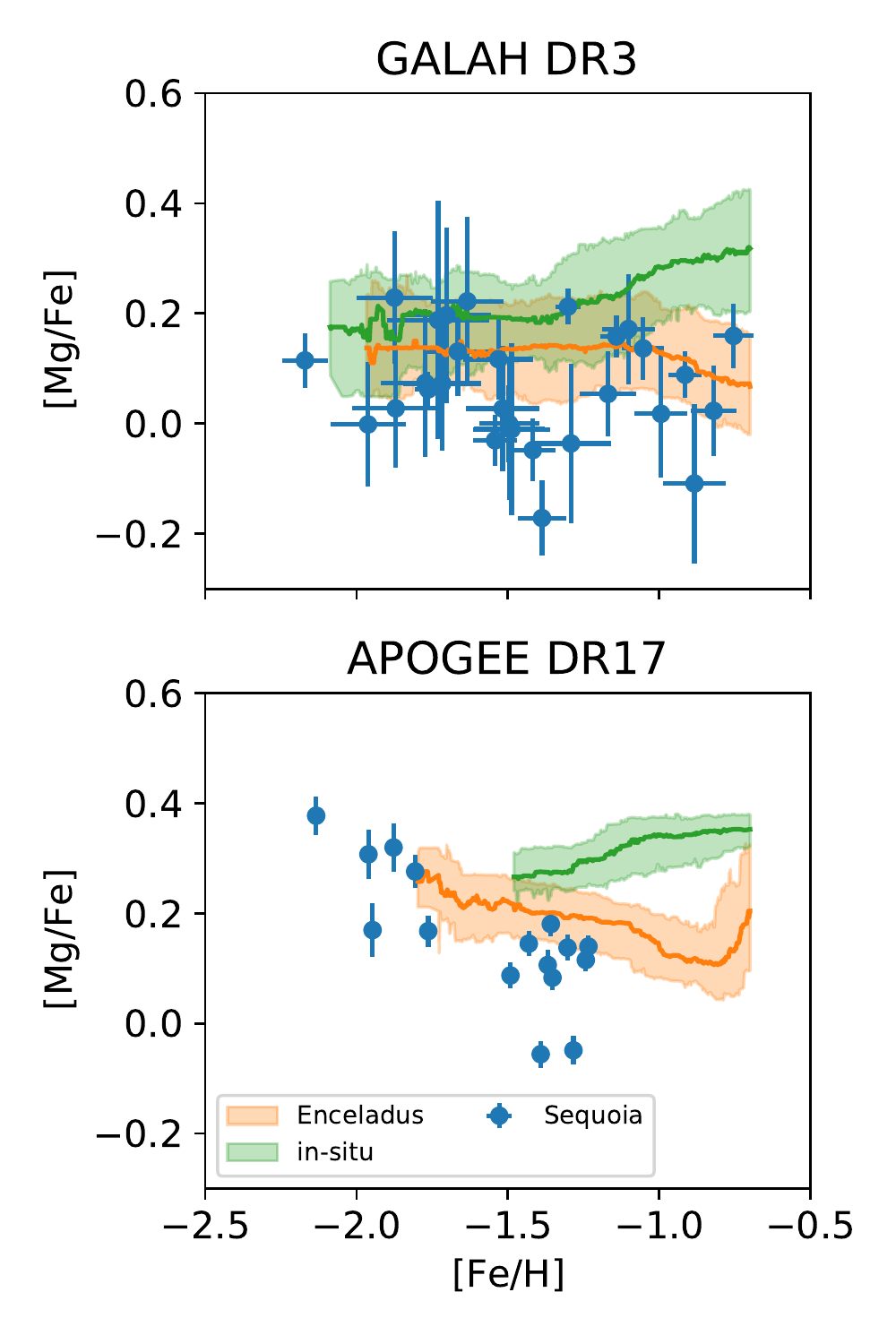}
\caption{Mg abundances of Sequoia, in situ stars, and Enceladus stars from GALAH DR3 and APOGEE DR17. The abundances of in situ stars and Enceladus stars are shown with a running median with a bin size of $0.1\,\mathrm{dex}$. The shaded regions show the 16-84 percentile region. \label{fig:surveyMg}}
\end{figure}

In this section, we compare our results with those seen in previous studies and in surveys.  
As we have discussed in Section~\ref{sec:intro}, previous studies do not agree on the chemical properties of Sequoia \citep{Matsuno2019a,Koppelman2019a,Monty2020a,Aguado2021a,Feuillet2020a}.
We revisit this problem using the updated data from spectroscopic surveys and a similar kinematic selection as we used in the present study. 

We selected Sequoia member candidates following that used in \citet{Koppelman2019a}.
The upper limits on circularity and energy were changed to $-0.35$ from $-0.40$ and to $-0.9\times 10^5\,\mathrm{km^2s^{-2}}$ from $-1.0\times 10^5\,\mathrm{km^2s^{-2}}$ so that the selection covers the kinematic extent of the stars in the present study. 

In Figure~\ref{fig:seq_lite}, we plotted Sequoia candidates from \citet{Stephens2002}, \citet{Ishigaki2012}, and \citet{Reggiani2018a}, which contain more than one Sequoia candidate at $-2<[\mathrm{Fe/H}]<-1$.
The abundance from \citet{Stephens2002} was updated following \citet{Monty2020a}.
There are nine stars from \citet{Stephens2002} that satisfy the kinematic selection for Sequoia, of which six are at $-2<[\mathrm{Fe/H}]<-1$.
The median reported uncertainty in [{Mg}/{Fe}] is $0.17\,\mathrm{dex}$, and hence it is difficult to chemically separate individual Sequoia stars (Section~\ref{sec:chemicalseparation}).
However, it is still possible to detect the average difference; this is why \citet{Venn2004a} were able to conclude that the most retrograde stars have a lower [{Mg}/{Fe}].
The situation with \citet{Ishigaki2012} is similar to that of \citet{Stephens2002} since the median uncertainty is $0.11\,\mathrm{dex}$ and the number of Sequoia stars is five (four in $-2<[\mathrm{Fe/H}]<-1$).

On the other hand, \citet{Reggiani2018a} conducted a very high precision abundance analysis for a pair of HD134439 and HD134440 stars (e.g., $\sigma([\mathrm{Mg/Fe}])\sim 0.02$).
With this precision, they are able to conclude that these two stars have a lower [$\alpha$/{Fe}] than the low-$\alpha$ population from NS10.  

Figure~\ref{fig:surveyMg} shows the Mg abundance of Sequoia candidates from GALAH DR3 and APOGEE DR17.
We selected APOGEE stars with $\mathtt{ASPCAPFLAG}=0$, $\mathtt{STARFLAG}=0$, $\mathtt{TEFF}<5500$, $\mathtt{LOGG}<3.5$, $\texttt{FE\_H\_FLAG}=0$, and $\texttt{MG\_FE\_FLAG}=0$. 
We also removed the known calibration cluster members and probable stellar cluster members.
For GALAH DR3, we selected stars with $\texttt{teff}<5500$, $\texttt{logg}<3.5$, $\texttt{flag\_fe\_h}=0$, and $\texttt{flag\_Mg\_fe}=0$.
We selected in situ stars and Gaia-Enceladus stars following \citet{Matsuno2021b} and computed the running median and 16 and 84 percentiles of [{Mg}/{Fe}] as a function of [{Fe}/{H}] using a $0.1\,\mathrm{dex}$ bin size. 
We note that APOGEE yields a smaller uncertainty at this metallicity simply because their targets are brighter and the target $S/N$ is higher than GALAH.

In both surveys, Sequoia stars clearly have a lower [{Mg}/{Fe}] than Gaia-Enceladus (Figure~\ref{fig:surveyMg}), reproducing our own finding, although the absolute values of [{Mg}/{Fe}] are different because of different approaches for abundance calculation and/or stellar parameter determination, for instance.
The low abundance of Mg for Sequoia was not clearly seen in previous studies \citep[e.g.,][]{Koppelman2019a,Myeong2019a,Feuillet2021a}, even though these authors used APOGEE data.
This is likely because of different selection criteria and the sample size.
The selection we made in the $L_z-E_n$ space allows for a smaller contamination compared to normalized action selections adopted in \citet{Myeong2019a} and \citet{Feuillet2021a} which includes low-$E_n$ stars
\footnote{The normalized action space is a space defined by $J_\phi/J_{\rm tot}$ and $(J_z-J_R)/J_{\rm tot}$, where $J_\phi,\,J_z, \,$ and $J_R$ are the azimuthal, vertical, and radial actions, respectively, and $J_{\rm tot}=|J_{\phi}|+J_z+J_R$. Although a selection in this normalized action space includes both stars at high $E_n$ and low $E_n$, it should in principle  be possible to make an action-based selection that is equivalent to the $L_z-E_n$ selection.}.
In order to deposit stars to a wide orbital energy range, the progenitor needs to be much more massive \citep{Koppelman2019a}, which does not seem to be the case for Sequoia.
The sample size with reliable abundance and reliable astrometric measurements has increased compared to more recent data releases from Gaia and APOGEE. 

We also see an abundance difference in other elements including Al (APOGEE) and K (GALAH).
However, a detailed investigation of Sequoia stars in large surveys is beyond the scope of the present study and is reserved for future studies. 

We finally note that both surveys seem to cover a wider metallicity range than the present study. 
The narrow metallicity range of our study could be due to the metallicity cut at $[\mathrm{Fe/H}]<-1$ for stars from LAMOST and the removal of two Sequoia candidates at a low metallicity as described in Section~\ref{sec:obs}.

\section{Conclusion\label{sec:conclusion}}

Through a differential abundance analysis of high-S/N and high-resolution spectra, we have shown that Sequoia stars are chemically distinguishable from Gaia-Enceladus.
The eight Sequoia stars in the metallicity range of $-1.8<[\mathrm{Fe/H}]<-1.4$ have a lower [{Na}/{Fe}], [{Mg}/{Fe}], [{Ca}/{Fe}], [{Ti}/{Fe}], [{Zn}/{Fe}], and [{Y}/{Fe}] compared to the values expected for Gaia-Enceladus. 
The abundance difference is $\sim 0.2\,\mathrm{dex}$ in [{Na}/{Fe}] and in [{Mg}/{Fe}] and $\sim 0.1\,\mathrm{dex}$ in other abundance ratios. 
This pattern in the abundance difference is similar to that between Gaia-Enceladus and in situ stars at a higher metallicity.
This suggests that Sequoia started experiencing chemical enrichment from SNe~Ia at a lower metallicity than Gaia-Enceladus.
We, however, note that we do not see a significant difference between Sequoia and Gaia-Enceladus in Ni abundance unlike in the comparison between Gaia-Enceladus and in situ stars, which might suggest that dominant types of SNe~Ia are different between Sequoia and Gaia-Enceladus.

We have also shown that Sequoia stars show low [{Y}/{Ba}] ratios, although its cause remains unclear.
We will provide abundances for additional neutron-capture elements (e.g., Sr and Eu) in a future study to separate the contribution of a weak $s$ process, main $s$ process, and $r$ process.  

We have further shown that separation in Sequoia and Gaia-Enceladus becomes most prominent when we take the average of [{Na}/{Fe}], [{Mg}/{Fe}], and [{Ca}/{Fe}], although this choice could vary depending on the data set.
We have shown that individual Sequoia stars can be chemically separated if the abundance precision in [{X}/{Fe}] is better than $0.07\,\mathrm{dex}$.
On the contrary, detecting the average abundance difference is much easier since the uncertainty on the mean scales with the square root of the number of stars if a kinematic selection that minimizes the contamination is adopted, and if there are few contaminants and chemical outliers.

Using the average of [{Na}/{Fe}], [{Mg}/{Fe}], and [{Ca}/{Fe}], we have concluded that eight out of the 12 stars we studied have distinct chemical abundances compared to Gaia-Enceladus.
These eight stars are most likely true members of Sequoia.
Only two of the remaining four stars seem to be contaminants from Gaia-Enceladus, indicating that the kinematic selection we adopted efficiently selects Sequoia stars.
For the remaining two stars, it is not clear if they have the same chemical abundance as the other Sequoia stars or if they are contaminants from Gaia-Enceladus because of their low metallicity and/or insufficient precision in our abundance measurements.

We have demonstrated that we can see kinematically selected Sequoia stars having lower Na, Mg, and Ca abundances also in data from the literature \citep{Stephens2002,Ishigaki2012,Reggiani2018a}.
\citet{Reggiani2018a} provide a sufficiently precise chemical abundance for the pair of HD134439 and HD134440 to chemically associate them to Sequoia.
We also confirmed low Mg abundances of Sequoia stars using GALAH DR3 and APOGEE DR17.

Now that we have established the chemical distinctness of Sequoia from the major populations in the halo, namely Gaia-Enceladus and in situ stars, future studies of chemical abundances of Sequoia stars using large spectroscopic surveys are obvious next steps. 
A large sample is necessary to study if the group of stars referred to as Sequoia in the present study can be further separated into a few subgroups \citep{Naidu2020a,Loevdal2022a,RuizLara2022a}. 
It would also be of interest to study the kinematic extent of chemically selected Sequoia stars. 
Large surveys that measure the chemical abundance of stars with a high-precision are necessary for these studies.

\begin{acknowledgements}

We thank Henrique Reggiani for providing the data that allow us to compare our results of equivalent width measurements and for checking their results in detail.
We also thank Xiaodi Yu and Ian Roederer for taking the high-resolution spectrum with MIKE on Magellan. 
Poul Erik Nissen provided useful comments, which prompted us to conduct an additional validation of our results.
We also thank the review made by the anonymous referee, which has improved the clarity of the manuscript.
This research has been supported by a Spinoza Grant from the Dutch Research Council (NWO).
WA, MNI, and TS were supported by JSPS KAKENHI Grant Number 21H04499.
This research is based in part on data collected at Subaru Telescope, which is operated by the National Astronomical Observatory of Japan.
We are honored and grateful for the opportunity of observing the Universe from Maunakea, which has the cultural, historical and natural significance in Hawaii.
This research has made use of the Keck Observatory Archive (KOA), which is operated by the W. M. Keck Observatory and the NASA Exoplanet Science Institute (NExScI), under contract with the National Aeronautics and Space Administration.
Part of the data were retrieved from the JVO portal (http://jvo.nao.ac.jp/portal/) operated by ADC/NAOJ.
This work is partly based on data obtained from the ESO Science Archive Facility, which are based on observations collected at the European Organisation for Astronomical Research in the Southern Hemisphere under ESO programmes 67.D-0086(A), 95.D-0504(A), 095.D-0504(A).

\end{acknowledgements}

\begin{appendix}
\section{Analysis of Gaia EDR3 360456543361799808\label{appendixA}}
Gaia EDR3 360456543361799808 turned out not to belong to Sequoia after updating the radial velocity to $-377.6\,\mathrm{km\,s^{-1}}$ using our high-resolution spectroscopy from the LAMOST DR4 value of $-288.15\,\mathrm{km\,s^{-1}}$.
With this updated radial velocity, we estimated $L_z=-841\,\mathrm{kpc\,km\,s^{-1}}$ and $E_n=-1.580\times 10^5\,\mathrm{km^2\,s^{-2}}$.
We note that LAMOST DR6 no longer provides a radial velocity for this object and Gaia DR2 gives $-376.3,\mathrm{km\,s^{-1}}$, which is closer to our measurement from the high-resolution spectrum.

We still measured stellar parameters and abundances for this object.
The derived stellar parameters are $T_{\rm eff}=6172\pm 70\,\mathrm{K}$, $\log g=3.905\pm0.032$, $v_t=1.567\pm0.108\,\mathrm{km\,s^{-1}}$, and $[\mathrm{Fe/H}]=-1.865\pm0.019$.
The derived abundances are summarized in Table~\ref{tab:0360}.
This star has a comparable abundance to Gaia-Enceladus and in situ stars.

\begin{table}
\caption{Abundances of Gaia EDR3 360456543361799808 \label{tab:0360}}
\begin{tabular}{l*{5}{r}}\hline\hline
     & $N$ & [{X}/{H}] & $\sigma$ & [{X}/{Fe}] & $\sigma$\\\hline
FeI  &     86&    -1.809&     0.032&       ...&       ...\\
FeII &     14&    -1.851&     0.019&       ...&       ...\\
NaI  &      2&    -2.027&     0.092&    -0.218&     0.087\\
MgI  &      6&    -1.588&     0.039&     0.221&     0.039\\
SiI  &      1&    -1.432&     0.063&     0.377&     0.066\\
CaI  &     18&    -1.414&     0.032&     0.395&     0.032\\
TiI  &     12&    -1.354&     0.064&     0.455&     0.054\\
TiII &     10&    -1.442&     0.033&     0.410&     0.030\\
CrI  &      2&    -1.910&     0.089&    -0.101&     0.083\\
MnI  &      2&    -2.184&     0.057&    -0.375&     0.048\\
NiI  &     11&    -1.761&     0.049&     0.048&     0.045\\
ZnI  &      2&    -1.670&     0.045&     0.139&     0.041\\
YII  &      2&    -1.911&     0.049&    -0.060&     0.046\\
BaII &      3&    -2.032&     0.053&    -0.181&     0.049\\\hline
\end{tabular}
\end{table}

\section{Spectroscopic gravity for 2657\_5888 and 4587\_5696 \label{sec:appendixC}}

The two stars, 2657\_5888 and 4587\_5696, are not on metal-poor isochrones (Figure~\ref{fig:kinematicsCMD}) and have significantly different iron abundances from neutral and ionized species (Table~\ref{tab:abundance}).
We confirmed the offsets with different isochrones (BaSTI and MIST isochrones). 
These indicate that the Gaia EDR3 parallax and photometry might not be yielding correct $\log g$. 
An alternative method for the $\log g$ determination is by requiring ionization balance, namely by enforcing neutral and ionized iron lines to yield consistent abundances.

The $\log g$ values that realize ionization balance are 4.69 and 4.96 for 2657\_5888 and 4587\_5696, respectively. 
Table~\ref{tab:deltaA_IB} shows the difference between the abundances derived with these $\log g$ values and those obtained assuming parameters from Table~\ref{tab:parameters}.

The change in $\log g$ does not affect our conclusion.
Although the difference shown in Table~\ref{tab:parameters} is in [{X}/{H}], a smaller difference is usually found when the comparison is made in [{X}/{Fe}] since we used the Fe abundance derived from iron lines in the same ionization stage as the species X.
Our conclusions are mostly based on [{X}/{Fe}].
In addition, these changes in $\log g$ affect only two stars, one of which was not regarded as chemically compatible with being a part of Sequoia.

Nonetheless, it is worthwhile mentioning how elemental abundances are affected.
For example, the change in [{X}/{Fe}] is negative in Mg, Ca, Ti~I, Ti~II, Y~II, and Ba~II; additionally, [{Mg}/{Fe}], [{Ca}/{Fe}], [{Ti}/{Fe}], [{Y}/{Fe}], and [{Ba}/{Fe}] of the two stars might be lower than shown in Figures~\ref{fig:Mg}, \ref{fig:alpha}, and \ref{fig:ncapture}.
On the other hand, [{Si}/{Fe}] and [{Zn}/{Fe}] ratios would be shifted higher if we adopted $\log g$ from the ionization balance. 
Effects on [{X}/{Fe}] of Cr, Mn, and Ni are minimal and the effect on [{Na}/{Fe}] is quite different between the two stars.

\begin{table}
  \caption{Abundance difference due to different $\log g$ \label{tab:deltaA_IB}}
  \centering
  \begin{tabular}{lrr}
\hline\hline
                              & 2657\_5888 & 4587\_5696 \\ \hline
$\Delta [\mathrm{Fe/H}]_{\rm I }$ &  -0.036    & 0.018      \\  
$\Delta [\mathrm{Na/H}]_{\rm I }$ &  0.089     & -0.005     \\  
$\Delta [\mathrm{Mg/H}]_{\rm I }$ &  -0.050    & -0.044     \\  
$\Delta [\mathrm{Si/H}]_{\rm I }$ &  0.058     & 0.076      \\  
$\Delta [\mathrm{Ca/H}]_{\rm I }$ & -0.106     & -0.071     \\  
$\Delta [\mathrm{Ti/H}]_{\rm I }$ & -0.109     & -0.016     \\  
$\Delta [\mathrm{Cr/H}]_{\rm I }$ & -0.044     & -0.040     \\  
$\Delta [\mathrm{Mn/H}]_{\rm I }$ & -0.052     &  0.013     \\  
$\Delta [\mathrm{Ni/H}]_{\rm I }$ &  0.011     &  0.050     \\  
$\Delta [\mathrm{Zn/H}]_{\rm I }$ &  0.099     &  0.142     \\  
$\Delta [\mathrm{Fe/H}]_{\rm II}$ &  0.178     &  0.202     \\  
$\Delta [\mathrm{Ti/H}]_{\rm II}$ &  0.121     &  0.165     \\  
$\Delta [\mathrm{Y/H}]_{\rm II}$ &  0.149     &  0.200     \\  
$\Delta [\mathrm{Ba/H}]_{\rm II}$ &  0.113     &  0.154     \\ \hline 
  \end{tabular}
\end{table}

Possible reasons for the offset of the two stars on the color-magnitude diagram are as follows: i) they are parts of unresolved binaries; ii) their extinctions are under-estimated; iii) Gaia does not provide correct astrometry and/or photometry; and iv) they experienced spacial process by which their current positions on the color-magnitude diagram does not reflect their age and metallicity.
Below, we discuss these four possibilities.

The first possibility seems most likely.
Although there are no signatures of the presence of companions in our high-resolution spectra, it is possible that the maximum velocity difference is too small to see the binary signature in our high-resolution spectra (smaller than a few $\mathrm{km\,s^{-1}}$) and, at the same time, the on-sky separation is smaller than what Gaia or photometric surveys can spatially resolve (less than a few tenths of an arcsec).
In addition, there is a possibility that we are not seeing any signatures of binarity simply because the orbit is at the phase where the radial velocity difference between the components is small. Long-term monitoring of the variation in high-resolution spectra will be able to provide an answer to this.
We also note that the $\texttt{ruwe}$ value in Gaia EDR3, which can be significantly higher than one because of astrometric jitter caused by unresolved companions, is $\sim 1.73$ for 4587\_5696, supporting this scenario.

We can rule out the second possibility since there is no sign of significant extinction in the high-resolution spectra in the form of Na~I interstellar absorptions.
We will be able to test the third possibility with improved astrometry and photometry from future Gaia data releases.
If all the other scenarios do not hold with future Gaia data, we can then examine the fourth possibility.

\section{Uncertainties in stellar parameters\label{app:uncertainty}}

Stellar parameters are dependent on each other and hence are determined iteratively.
We, therefore, need to take the uncertainties in the other parameters into account when estimating the uncertainty in one parameter.
We also consider correlations between stellar parameters and their effects on abundances.
The goal of this section is to obtain the covariance matrix $\Sigma$ among the four stellar parameters.

We consider four stellar parameters, $x_1=T_{\rm eff},\,x_2=\log g,\,x_3=v_t,\,$and $x_4=[\mathrm{Fe/H}]_{\rm sp}$.
A parameter $x_i$ was determined through a function $f_i(\mathbf{x})$; the set of best estimates $\tilde{\mathbf{x}}$ satisfies $\tilde{x}_i=f_i(\tilde{\mathbf{x}})$.
We first estimated the uncertainty in each parameter $\epsilon_i$ by fixing other parameters as described in Section~\ref{sec:parameters}.
The $\epsilon_i$ is not necessarily close to realistic uncertainty since it neglects the effect of the uncertainties in other parameters.
The values we estimated can be expressed as
\begin{equation}
x_i = f_i(\textbf{x}) + \epsilon_i, \label{eq:app1}
\end{equation}
which can be approximated as
\begin{equation}
x_i \simeq f_i(\tilde{\mathbf{x}})+\sum_{i\neq j} \frac{\partial f_i}{\partial x_j}(x_j-\tilde{x}_{j})+\epsilon_i.
\end{equation}
Here we define $\delta \mathbf{x}=\mathbf{x}-\mathbf{\tilde{x}}$ and the matrix $\mathbf{A}$ whose element is 
\begin{equation}
A_{ij}=
\begin{cases}
0 & (i=j)\\
\frac{\partial f_i}{\partial x_j} &(i\neq j). 
\end{cases}
\end{equation}
From equation \ref{eq:app1}, we can write 
\begin{equation}
\delta\textbf{x}=\epsilon+\textbf{A}\delta\textbf{x},
\end{equation}
hence 
\begin{equation}
(\textbf{I}-\textbf{A})\delta\textbf{x}=\epsilon.
\end{equation}
Since $\Sigma = \langle\delta\textbf{x}\delta\textbf{x}^T\rangle$ and $\langle\epsilon_i,\epsilon_j\rangle=\delta_{ij}\epsilon_i^2$, the covariance matrix can be calculated from 
\begin{equation}
\Sigma=(\textbf{I}-\textbf{A})^{-1}\textrm{diag}(\epsilon_i^2)[(\textbf{I}-\textbf{A})^{-1}]^T. \label{eq:sigma}
\end{equation}

The above calculation is equivalent to considering the following likelihood,
\begin{equation}
{\cal L} \propto \prod \exp[-\frac{1}{2}\frac{(x_i-f_i(\mathbf{x}))^2}{\epsilon_i^2}],
\end{equation}
and calculating Fisher's matrix ${\cal F}$, whose element is expressed as 
\begin{equation}
{\cal F}_{lm} = -\frac{\partial^2 \log {\cal L}}{\partial x_l \partial x_m}.
\end{equation}
Assuming $\textbf{x}$ follows a multivariate Gaussian distribution, ${\cal F}$ is equal to $\Sigma^{-1}$ \citep[e.g.,][]{Andrae2010a}.
Eq. \ref{eq:sigma} can also be derived from this equation.

In practice, we estimated $A_{ij}$ by redetermining the parameter $i$ while shifting the parameter $j$ by $\pm\epsilon_j$.
Since we determined both $T_{\rm eff}$ and $v_t$ from neutral Fe lines, the correlation between these two parameters can be significant.

\section{Additional table}

\clearpage
\onecolumn

\begin{landscape}
\begin{longtable}{lrrrrrrrrrrrrrrrrr}
\caption{\label{tab:abundance}Abundances of Sequoia stars}\\
\hline\hline
\endfirsthead
\caption{continued.}\\
\hline\hline
\endhead
\hline
\endfoot
     &                     \multicolumn{5}{c}{1336\_6432}&&                     \multicolumn{5}{c}{2657\_5888}&&                     \multicolumn{5}{c}{2813\_6032}\\\cline{2-6}\cline{8-12}\cline{14-18}
     & $N$ & [{X}/{H}] & $\sigma$ & [{X}/{Fe}] & $\sigma$&& $N$ & [{X}/{H}] & $\sigma$ & [{X}/{Fe}] & $\sigma$&& $N$ & [{X}/{H}] & $\sigma$ & [{X}/{Fe}] & $\sigma$\\\hline
FeI  &     84&    -1.671&     0.033&       ...&       ...&&     90&    -1.343&     0.048&       ...&       ...&&     97&    -1.602&     0.030&       ...&       ...\\
FeII &     10&    -1.692&     0.027&       ...&       ...&&      9&    -1.517&     0.045&       ...&       ...&&     16&    -1.632&     0.026&       ...&       ...\\
NaI  &      2&    -2.079&     0.085&    -0.408&     0.081&&      4&    -1.522&     0.071&    -0.179&     0.076&&      4&    -1.936&     0.063&    -0.334&     0.062\\
MgI  &      6&    -1.563&     0.039&     0.108&     0.038&&      2&    -1.087&     0.086&     0.256&     0.082&&      6&    -1.461&     0.037&     0.141&     0.036\\
SiI  &      2&    -1.533&     0.064&     0.137&     0.065&&      4&    -1.025&     0.035&     0.318&     0.053&&      1&    -1.242&     0.061&     0.360&     0.063\\
CaI  &     15&    -1.354&     0.034&     0.316&     0.032&&     13&    -0.947&     0.067&     0.396&     0.062&&     19&    -1.361&     0.032&     0.241&     0.030\\
TiI  &      9&    -1.293&     0.056&     0.378&     0.044&&     16&    -1.110&     0.113&     0.232&     0.102&&     10&    -1.272&     0.053&     0.330&     0.042\\
TiII &     10&    -1.421&     0.035&     0.271&     0.034&&      8&    -1.273&     0.059&     0.245&     0.064&&     10&    -1.403&     0.034&     0.230&     0.034\\
CrI  &      3&    -1.755&     0.066&    -0.084&     0.055&&      3&    -1.343&     0.134&    -0.001&     0.123&&      3&    -1.640&     0.061&    -0.039&     0.052\\
MnI  &      2&    -1.968&     0.055&    -0.297&     0.046&&      2&    -1.679&     0.114&    -0.336&     0.103&&      3&    -1.912&     0.048&    -0.310&     0.040\\
NiI  &      8&    -1.711&     0.047&    -0.041&     0.042&&     22&    -1.464&     0.057&    -0.121&     0.063&&     15&    -1.659&     0.040&    -0.057&     0.036\\
ZnI  &      2&    -1.654&     0.048&     0.017&     0.043&&      2&    -1.354&     0.039&    -0.011&     0.060&&      2&    -1.534&     0.045&     0.068&     0.041\\
YII  &      1&    -1.828&     0.061&    -0.136&     0.060&&      1&    -1.695&     0.062&    -0.178&     0.080&&      1&    -1.745&     0.054&    -0.113&     0.053\\
BaII &      4&    -2.074&     0.052&    -0.382&     0.049&&      3&    -1.709&     0.050&    -0.192&     0.063&&      4&    -2.021&     0.057&    -0.388&     0.056\\\hline\hline
     &                     \multicolumn{5}{c}{2870\_9072}&&                     \multicolumn{5}{c}{3336\_0672}&&                     \multicolumn{5}{c}{3341\_2720}\\\cline{2-6}\cline{8-12}\cline{14-18}
     & $N$ & [{X}/{H}] & $\sigma$ & [{X}/{Fe}] & $\sigma$&& $N$ & [{X}/{H}] & $\sigma$ & [{X}/{Fe}] & $\sigma$&& $N$ & [{X}/{H}] & $\sigma$ & [{X}/{Fe}] & $\sigma$\\\hline
FeI  &    104&    -1.451&     0.025&       ...&       ...&&     94&    -1.702&     0.035&       ...&       ...&&     79&    -1.742&     0.031&       ...&       ...\\
FeII &     17&    -1.438&     0.024&       ...&       ...&&     16&    -1.667&     0.028&       ...&       ...&&     10&    -1.822&     0.029&       ...&       ...\\
NaI  &      3&    -1.919&     0.055&    -0.468&     0.054&&      3&    -2.012&     0.163&    -0.310&     0.160&&      2&    -2.162&     0.095&    -0.420&     0.090\\
MgI  &      7&    -1.455&     0.039&    -0.004&     0.039&&      7&    -1.622&     0.072&     0.079&     0.074&&      6&    -1.626&     0.040&     0.116&     0.039\\
SiI  &      5&    -1.285&     0.042&     0.166&     0.044&&      1&    -1.264&     0.068&     0.438&     0.070&&      1&    -1.152&     0.080&     0.590&     0.082\\
CaI  &     16&    -1.291&     0.035&     0.160&     0.033&&     20&    -1.425&     0.061&     0.277&     0.061&&     21&    -1.422&     0.035&     0.320&     0.033\\
TiI  &     12&    -1.357&     0.058&     0.094&     0.052&&     14&    -1.586&     0.082&     0.116&     0.080&&      9&    -1.321&     0.070&     0.421&     0.061\\
TiII &     11&    -1.384&     0.038&     0.054&     0.033&&      9&    -1.504&     0.032&     0.163&     0.037&&     11&    -1.492&     0.051&     0.331&     0.046\\
CrI  &      5&    -1.477&     0.058&    -0.026&     0.052&&      4&    -1.730&     0.127&    -0.028&     0.121&&      2&    -1.725&     0.094&     0.017&     0.088\\
MnI  &      3&    -1.882&     0.050&    -0.431&     0.045&&      4&    -2.151&     0.109&    -0.449&     0.105&&      2&    -2.001&     0.055&    -0.259&     0.047\\
NiI  &     22&    -1.564&     0.036&    -0.113&     0.034&&     17&    -1.728&     0.076&    -0.026&     0.075&&      4&    -1.786&     0.044&    -0.044&     0.039\\
ZnI  &      2&    -1.530&     0.039&    -0.079&     0.038&&      2&    -1.710&     0.048&    -0.009&     0.053&&      2&    -1.624&     0.058&     0.118&     0.054\\
YII  &      2&    -1.913&     0.048&    -0.475&     0.044&&      2&    -2.108&     0.092&    -0.441&     0.096&&      2&    -1.824&     0.058&    -0.002&     0.053\\
BaII &      4&    -1.879&     0.050&    -0.441&     0.044&&      4&    -2.005&     0.050&    -0.338&     0.054&&      4&    -2.114&     0.061&    -0.292&     0.053\\\hline\hline
     &                     \multicolumn{5}{c}{4587\_5616}&&                     \multicolumn{5}{c}{4850\_5696}&&                     \multicolumn{5}{c}{G115-58   }\\\cline{2-6}\cline{8-12}\cline{14-18}
     & $N$ & [{X}/{H}] & $\sigma$ & [{X}/{Fe}] & $\sigma$&& $N$ & [{X}/{H}] & $\sigma$ & [{X}/{Fe}] & $\sigma$&& $N$ & [{X}/{H}] & $\sigma$ & [{X}/{Fe}] & $\sigma$\\\hline
FeI  &     93&    -1.563&     0.039&       ...&       ...&&     98&    -1.684&     0.026&       ...&       ...&&    106&    -1.317&     0.029&       ...&       ...\\
FeII &      6&    -1.730&     0.053&       ...&       ...&&     13&    -1.708&     0.027&       ...&       ...&&     18&    -1.364&     0.021&       ...&       ...\\
NaI  &      4&    -1.964&     0.058&    -0.400&     0.056&&      3&    -2.095&     0.066&    -0.411&     0.064&&      3&    -1.736&     0.062&    -0.418&     0.059\\
MgI  &      6&    -1.551&     0.056&     0.013&     0.054&&      4&    -1.574&     0.047&     0.111&     0.045&&      5&    -1.278&     0.050&     0.039&     0.047\\
SiI  &      2&    -1.242&     0.147&     0.322&     0.150&&      1&    -1.583&     0.079&     0.101&     0.079&&      3&    -1.175&     0.060&     0.142&     0.062\\
CaI  &     13&    -1.376&     0.052&     0.187&     0.047&&     23&    -1.473&     0.032&     0.211&     0.030&&     19&    -1.085&     0.035&     0.232&     0.031\\
TiI  &     15&    -1.403&     0.079&     0.161&     0.067&&     10&    -1.366&     0.055&     0.318&     0.047&&     10&    -1.149&     0.062&     0.168&     0.052\\
TiII &     10&    -1.617&     0.066&     0.112&     0.078&&     14&    -1.540&     0.037&     0.167&     0.039&&      9&    -1.106&     0.059&     0.258&     0.057\\
CrI  &      5&    -1.463&     0.082&     0.101&     0.071&&      6&    -1.638&     0.085&     0.046&     0.080&&      4&    -1.309&     0.060&     0.008&     0.051\\
MnI  &      4&    -1.984&     0.090&    -0.420&     0.083&&      8&    -2.002&     0.061&    -0.318&     0.055&&      3&    -1.634&     0.052&    -0.316&     0.045\\
NiI  &     18&    -1.663&     0.053&    -0.100&     0.051&&      9&    -1.770&     0.052&    -0.086&     0.049&&     14&    -1.440&     0.043&    -0.123&     0.038\\
ZnI  &      2&    -1.570&     0.035&    -0.007&     0.048&&      2&    -1.605&     0.043&     0.079&     0.039&&      2&    -1.411&     0.044&    -0.094&     0.040\\
YII  &      2&    -1.781&     0.051&    -0.051&     0.075&&      1&    -2.055&     0.058&    -0.347&     0.060&&      2&    -1.567&     0.045&    -0.204&     0.041\\
BaII &      4&    -1.842&     0.052&    -0.113&     0.071&&      4&    -2.181&     0.054&    -0.473&     0.055&&      4&    -1.518&     0.049&    -0.154&     0.045\\\hline\hline
     &                     \multicolumn{5}{c}{G90-36    }&&                     \multicolumn{5}{c}{HIP28104  }&&                     \multicolumn{5}{c}{HIP98492  }\\\cline{2-6}\cline{8-12}\cline{14-18}
     & $N$ & [{X}/{H}] & $\sigma$ & [{X}/{Fe}] & $\sigma$&& $N$ & [{X}/{H}] & $\sigma$ & [{X}/{Fe}] & $\sigma$&& $N$ & [{X}/{H}] & $\sigma$ & [{X}/{Fe}] & $\sigma$\\\hline
FeI  &     80&    -1.633&     0.023&       ...&       ...&&    101&    -1.973&     0.021&       ...&       ...&&    103&    -1.291&     0.021&       ...&       ...\\
FeII &      6&    -1.622&     0.040&       ...&       ...&&     15&    -1.980&     0.025&       ...&       ...&&     18&    -1.246&     0.022&       ...&       ...\\
NaI  &      1&    -2.106&     0.081&    -0.473&     0.081&&      2&    -2.332&     0.071&    -0.358&     0.070&&      4&    -1.429&     0.063&    -0.138&     0.062\\
MgI  &      4&    -1.565&     0.043&     0.069&     0.040&&      6&    -1.812&     0.034&     0.161&     0.032&&      5&    -1.074&     0.040&     0.217&     0.037\\
SiI  &      2&    -1.289&     0.050&     0.345&     0.054&&      0&       ...&       ...&       ...&       ...&&      8&    -1.052&     0.042&     0.240&     0.044\\
CaI  &     16&    -1.479&     0.035&     0.154&     0.032&&     22&    -1.651&     0.025&     0.322&     0.023&&     16&    -1.048&     0.031&     0.243&     0.028\\
TiI  &     13&    -1.585&     0.052&     0.048&     0.044&&      9&    -1.544&     0.042&     0.429&     0.034&&     16&    -1.157&     0.042&     0.134&     0.036\\
TiII &      4&    -1.552&     0.045&     0.070&     0.052&&     14&    -1.723&     0.030&     0.257&     0.030&&      7&    -0.959&     0.041&     0.287&     0.037\\
CrI  &      4&    -1.651&     0.055&    -0.018&     0.048&&      4&    -1.944&     0.048&     0.029&     0.043&&      4&    -1.283&     0.049&     0.008&     0.043\\
MnI  &      3&    -2.019&     0.063&    -0.386&     0.059&&      6&    -2.357&     0.045&    -0.384&     0.039&&      4&    -1.688&     0.083&    -0.397&     0.081\\
NiI  &     17&    -1.730&     0.042&    -0.097&     0.040&&      8&    -2.018&     0.059&    -0.045&     0.058&&     24&    -1.302&     0.026&    -0.011&     0.025\\
ZnI  &      2&    -1.672&     0.043&    -0.039&     0.048&&      1&    -1.858&     0.059&     0.116&     0.057&&      2&    -1.083&     0.036&     0.208&     0.039\\
YII  &      1&    -1.900&     0.063&    -0.278&     0.072&&      1&    -2.038&     0.079&    -0.058&     0.079&&      2&    -1.506&     0.037&    -0.260&     0.034\\
BaII &      4&    -1.876&     0.043&    -0.254&     0.052&&      4&    -2.332&     0.043&    -0.352&     0.044&&      4&    -1.612&     0.043&    -0.366&     0.040
\end{longtable}
\end{landscape}

\end{appendix}


\begin{thebibliography}{92}
\expandafter\ifx\csname natexlab\endcsname\relax\def\natexlab#1{#1}\fi

\bibitem[{Aguado {et~al.}(2021)Aguado, Belokurov, Myeong, Evans, Kobayashi,
  Sbordone, Chanam{\'e}, Navarrete, \& Koposov}]{Aguado2021a}
Aguado, D.~S., Belokurov, V., Myeong, G.~C., {et~al.} 2021, \apjl, 908, L8

\bibitem[{Andrae(2010)}]{Andrae2010a}
Andrae, R. 2010, arXiv e-prints, arXiv:1009.2755

\bibitem[{Antoja {et~al.}(2020)Antoja, Ramos, Mateu, Helmi, Anders, Jordi, \&
  Carballo-Bello}]{Antoja2020a}
Antoja, T., Ramos, P., Mateu, C., {et~al.} 2020, \aap, 635, L3

\bibitem[{{Belokurov} {et~al.}(2018){Belokurov}, {Erkal}, {Evans}, {Koposov},
  \& {Deason}}]{Belokurov2018a}
{Belokurov}, V., {Erkal}, D., {Evans}, N.~W., {Koposov}, S.~E., \& {Deason},
  A.~J. 2018, \mnras, 478, 611

\bibitem[{Belokurov {et~al.}(2020)Belokurov, Sanders, Fattahi, Smith, Deason,
  Evans, \& Grand}]{Belokurov2020a}
Belokurov, V., Sanders, J.~L., Fattahi, A., {et~al.} 2020, \mnras, 494, 3880

\bibitem[{{Belokurov} {et~al.}(2006){Belokurov}, {Zucker}, {Evans}, {Gilmore},
  {Vidrih}, {Bramich}, {Newberg}, {Wyse}, {Irwin}, {Fellhauer}, {Hewett},
  {Walton}, {Wilkinson}, {Cole}, {Yanny}, {Rockosi}, {Beers}, {Bell},
  {Brinkmann}, {Ivezi{\'c}}, \& {Lupton}}]{Belokurov2006a}
{Belokurov}, V., {Zucker}, D.~B., {Evans}, N.~W., {et~al.} 2006, \apjl, 642,
  L137

\bibitem[{Bennett \& Bovy(2019)}]{Bennett2019a}
Bennett, M. \& Bovy, J. 2019, \mnras, 482, 1417

\bibitem[{{Bernard} {et~al.}(2016){Bernard}, {Ferguson}, {Schlafly}, {Martin},
  {Rix}, {Bell}, {Finkbeiner}, {Goldman}, {Mart{\'{\i}}nez-Delgado}, {Sesar},
  {Wyse}, {Burgett}, {Chambers}, {Draper}, {Hodapp}, {Kaiser}, {Kudritzki},
  {Magnier}, {Metcalfe}, {Wainscoat}, \& {Waters}}]{Bernard2016}
{Bernard}, E.~J., {Ferguson}, A.~M.~N., {Schlafly}, E.~F., {et~al.} 2016,
  \mnras, 463, 1759

\bibitem[{Bernstein {et~al.}(2003)Bernstein, Shectman, Gunnels, Mochnacki, \&
  Athey}]{Bernstein2003a}
Bernstein, R., Shectman, S.~A., Gunnels, S.~M., Mochnacki, S., \& Athey, A.~E.
  2003, in Society of Photo-Optical Instrumentation Engineers (SPIE) Conference
  Series, Vol. 4841, Instrument Design and Performance for Optical/Infrared
  Ground-based Telescopes, ed. M.~{Iye} \& A.~F.~M. {Moorwood}, 1694--1704

\bibitem[{Brook {et~al.}(2020)Brook, Kawata, Gibson, Gallart, \&
  Vicente}]{Brook2020a}
Brook, C.~B., Kawata, D., Gibson, B.~K., Gallart, C., \& Vicente, A. 2020,
  \mnras, 495, 2645

\bibitem[{Casagrande {et~al.}(2021)Casagrande, Lin, Rains, Liu, Buder, Horner,
  Asplund, Lewis, Martell, Nordlander, Stello, Ting, Wittenmyer,
  Bland-Hawthorn, Casey, De~Silva, D'Orazi, Freeman, Hayden, Kos, Lind,
  Schlesinger, Sharma, Simpson, Zucker, \& Zwitter}]{Casagrande2021a}
Casagrande, L., Lin, J., Rains, A.~D., {et~al.} 2021, \mnras, 507, 2684

\bibitem[{Casagrande \& VandenBerg(2018)}]{Casagrande2018a}
Casagrande, L. \& VandenBerg, D.~A. 2018, \mnras, 479, L102

\bibitem[{Cayrel(1988)}]{Cayrel1988a}
Cayrel, R. 1988, in The Impact of Very High S/N Spectroscopy on Stellar
  Physics, ed. G.~{Cayrel de Strobel} \& M.~{Spite}, Vol. 132, 345

\bibitem[{{Chen} {et~al.}(2014){Chen}, {Herwig}, {Denissenkov}, \&
  {Paxton}}]{Chen2014a}
{Chen}, M.~C., {Herwig}, F., {Denissenkov}, P.~A., \& {Paxton}, B. 2014,
  \mnras, 440, 1274

\bibitem[{Chen \& Zhao(2006)}]{Chen2006a}
Chen, Y.~Q. \& Zhao, G. 2006, \mnras, 370, 2091

\bibitem[{Cohen {et~al.}(2013)Cohen, Christlieb, Thompson, McWilliam, Shectman,
  Reimers, Wisotzki, \& Kirby}]{Cohen2013a}
Cohen, J.~G., Christlieb, N., Thompson, I., {et~al.} 2013, \apj, 778, 56

\bibitem[{Cohen \& Huang(2009)}]{Cohen2009a}
Cohen, J.~G. \& Huang, W. 2009, \apj, 701, 1053

\bibitem[{Dekker {et~al.}(2000)Dekker, D'Odorico, Kaufer, Delabre, \&
  Kotzlowski}]{Dekker2000a}
Dekker, H., D'Odorico, S., Kaufer, A., Delabre, B., \& Kotzlowski, H. 2000, in
  Society of Photo-Optical Instrumentation Engineers (SPIE) Conference Series,
  Vol. 4008, Optical and IR Telescope Instrumentation and Detectors, ed.
  M.~{Iye} \& A.~F. {Moorwood}, 534--545

\bibitem[{Di~Matteo {et~al.}(2019)Di~Matteo, Haywood, Lehnert, Katz,
  Khoperskov, Snaith, G{\'o}mez, \& Robichon}]{DiMatteo2019a}
Di~Matteo, P., Haywood, M., Lehnert, M.~D., {et~al.} 2019, \aap, 632, A4

\bibitem[{Fern{\'a}ndez-Alvar {et~al.}(2018)Fern{\'a}ndez-Alvar, Carigi,
  Schuster, Hayes, {\'A}vila-Vergara, Majewski, Allende~Prieto, Beers,
  S{\'a}nchez, Zamora, Garc{\'\i}a-Hern{\'a}ndez, Tang, Fern{\'a}ndez-Trincado,
  Tissera, Geisler, \& Villanova}]{FernandezAlvar2018a}
Fern{\'a}ndez-Alvar, E., Carigi, L., Schuster, W.~J., {et~al.} 2018, \apj, 852,
  50

\bibitem[{Fern{\'a}ndez-Alvar {et~al.}(2019)Fern{\'a}ndez-Alvar,
  Fern{\'a}ndez-Trincado, Moreno, Schuster, Carigi, Recio-Blanco, Beers,
  Chiappini, Anders, Santiago, Queiroz, P{\'e}rez-Villegas, Zamora,
  Garc{\'\i}a-Hern{\'a}ndez, \& Ortigoza-Urdaneta}]{FernandezAlvar2019a}
Fern{\'a}ndez-Alvar, E., Fern{\'a}ndez-Trincado, J.~G., Moreno, E., {et~al.}
  2019, \mnras, 487, 1462

\bibitem[{Feuillet {et~al.}(2020)Feuillet, Feltzing, Sahlholdt, \&
  Casagrande}]{Feuillet2020a}
Feuillet, D.~K., Feltzing, S., Sahlholdt, C.~L., \& Casagrande, L. 2020,
  \mnras, 497, 109

\bibitem[{Feuillet {et~al.}(2021)Feuillet, Sahlholdt, Feltzing, \&
  Casagrande}]{Feuillet2021a}
Feuillet, D.~K., Sahlholdt, C.~L., Feltzing, S., \& Casagrande, L. 2021,
  \mnras, 508, 1489

\bibitem[{{Foreman-Mackey} {et~al.}(2013){Foreman-Mackey}, {Hogg}, {Lang}, \&
  {Goodman}}]{emcee}
{Foreman-Mackey}, D., {Hogg}, D.~W., {Lang}, D., \& {Goodman}, J. 2013, \pasp,
  125, 306

\bibitem[{{Gaia Collaboration} {et~al.}(2018){Gaia Collaboration}, {Brown},
  {Vallenari}, {Prusti}, {de Bruijne}, {Babusiaux}, {Bailer-Jones}, {Biermann},
  {Evans}, {Eyer}, {Jansen}, {Jordi}, {Klioner}, {Lammers}, {Lindegren},
  {Luri}, {Mignard}, {Panem}, {Pourbaix}, {Randich}, {Sartoretti}, {Siddiqui},
  {Soubiran}, {van Leeuwen}, {Walton}, {Arenou}, {Bastian}, {Cropper},
  {Drimmel}, {Katz}, {Lattanzi}, {Bakker}, {Cacciari}, {Casta{\~n}eda},
  {Chaoul}, {Cheek}, {De Angeli}, {Fabricius}, {Guerra}, {Holl}, {Masana},
  {Messineo}, {Mowlavi}, {Nienartowicz}, {Panuzzo}, {Portell}, {Riello},
  {Seabroke}, {Tanga}, {Th{\'e}venin}, {Gracia-Abril}, {Comoretto},
  {Garcia-Reinaldos}, {Teyssier}, {Altmann}, {Andrae}, {Audard},
  {Bellas-Velidis}, {Benson}, {Berthier}, {Blomme}, {Burgess}, {Busso},
  {Carry}, {Cellino}, {Clementini}, {Clotet}, {Creevey}, {Davidson}, {De
  Ridder}, {Delchambre}, {Dell'Oro}, {Ducourant},
  {Fern{\'a}ndez-Hern{\'a}ndez}, {Fouesneau}, {Fr{\'e}mat}, {Galluccio},
  {Garc{\'\i}a-Torres}, {Gonz{\'a}lez-N{\'u}{\~n}ez}, {Gonz{\'a}lez-Vidal},
  {Gosset}, {Guy}, {Halbwachs}, {Hambly}, {Harrison}, {Hern{\'a}ndez},
  {Hestroffer}, {Hodgkin}, {Hutton}, {Jasniewicz}, {Jean-Antoine-Piccolo},
  {Jordan}, {Korn}, {Krone-Martins}, {Lanzafame}, {Lebzelter}, {L{\"o}ffler},
  {Manteiga}, {Marrese}, {Mart{\'\i}n-Fleitas}, {Moitinho}, {Mora}, {Muinonen},
  {Osinde}, {Pancino}, {Pauwels}, {Petit}, {Recio-Blanco}, {Richards},
  {Rimoldini}, {Robin}, {Sarro}, {Siopis}, {Smith}, {Sozzetti}, {S{\"u}veges},
  {Torra}, {van Reeven}, {Abbas}, {Abreu Aramburu}, {Accart}, {Aerts},
  {Altavilla}, {{\'A}lvarez}, {Alvarez}, {Alves}, {Anderson}, {Andrei},
  {Anglada Varela}, {Antiche}, {Antoja}, {Arcay}, {Astraatmadja}, {Bach},
  {Baker}, {Balaguer-N{\'u}{\~n}ez}, {Balm}, {Barache}, {Barata}, {Barbato},
  {Barblan}, {Barklem}, {Barrado}, {Barros}, {Barstow}, {Bartholom{\'e}
  Mu{\~n}oz}, {Bassilana}, {Becciani}, {Bellazzini}, {Berihuete}, {Bertone},
  {Bianchi}, {Bienaym{\'e}}, {Blanco-Cuaresma}, {Boch}, {Boeche}, {Bombrun},
  {Borrachero}, {Bossini}, {Bouquillon}, {Bourda}, {Bragaglia}, {Bramante},
  {Breddels}, {Bressan}, {Brouillet}, {Br{\"u}semeister}, {Brugaletta},
  {Bucciarelli}, {Burlacu}, {Busonero}, {Butkevich}, {Buzzi}, {Caffau},
  {Cancelliere}, {Cannizzaro}, {Cantat-Gaudin}, {Carballo}, {Carlucci},
  {Carrasco}, {Casamiquela}, {Castellani}, {Castro-Ginard}, {Charlot},
  {Chemin}, {Chiavassa}, {Cocozza}, {Costigan}, {Cowell}, {Crifo}, {Crosta},
  {Crowley}, {Cuypers}, {Dafonte}, {Damerdji}, {Dapergolas}, {David}, {David},
  {de Laverny}, {De Luise}, {De March}, {de Martino}, {de Souza}, {de Torres},
  {Debosscher}, {del Pozo}, {Delbo}, {Delgado}, {Delgado}, {Di Matteo},
  {Diakite}, {Diener}, {Distefano}, {Dolding}, {Drazinos}, {Dur{\'a}n},
  {Edvardsson}, {Enke}, {Eriksson}, {Esquej}, {Eynard Bontemps}, {Fabre},
  {Fabrizio}, {Faigler}, {Falc{\~a}o}, {Farr{\`a}s Casas}, {Federici},
  {Fedorets}, {Fernique}, {Figueras}, {Filippi}, {Findeisen}, {Fonti},
  {Fraile}, {Fraser}, {Fr{\'e}zouls}, {Gai}, {Galleti}, {Garabato},
  {Garc{\'\i}a-Sedano}, {Garofalo}, {Garralda}, {Gavel}, {Gavras}, {Gerssen},
  {Geyer}, {Giacobbe}, {Gilmore}, {Girona}, {Giuffrida}, {Glass}, {Gomes},
  {Granvik}, {Gueguen}, {Guerrier}, {Guiraud}, {Guti{\'e}rrez-S{\'a}nchez},
  {Haigron}, {Hatzidimitriou}, {Hauser}, {Haywood}, {Heiter}, {Helmi}, {Heu},
  {Hilger}, {Hobbs}, {Hofmann}, {Holland}, {Huckle}, {Hypki}, {Icardi},
  {Jan{\ss}en}, {Jevardat de Fombelle}, {Jonker}, {Juh{\'a}sz}, {Julbe},
  {Karampelas}, {Kewley}, {Klar}, {Kochoska}, {Kohley}, {Kolenberg},
  {Kontizas}, {Kontizas}, {Koposov}, {Kordopatis}, {Kostrzewa-Rutkowska},
  {Koubsky}, {Lambert}, {Lanza}, {Lasne}, {Lavigne}, {Le Fustec}, {Le
  Poncin-Lafitte}, {Lebreton}, {Leccia}, {Leclerc}, {Lecoeur-Taibi},
  {Lenhardt}, {Leroux}, {Liao}, {Licata}, {Lindstr{\o}m}, {Lister}, {Livanou},
  {Lobel}, {L{\'o}pez}, {Managau}, {Mann}, {Mantelet}, {Marchal}, {Marchant},
  {Marconi}, {Marinoni}, {Marschalk{\'o}}, {Marshall}, {Martino}, {Marton},
  {Mary}, {Massari}, {Matijevi{\v{c}}}, {Mazeh}, {McMillan}, {Messina},
  {Michalik}, {Millar}, {Molina}, {Molinaro}, {Moln{\'a}r}, {Montegriffo},
  {Mor}, {Morbidelli}, {Morel}, {Morris}, {Mulone}, {Muraveva}, {Musella},
  {Nelemans}, {Nicastro}, {Noval}, {O'Mullane}, {Ord{\'e}novic},
  {Ord{\'o}{\~n}ez-Blanco}, {Osborne}, {Pagani}, {Pagano}, {Pailler},
  {Palacin}, {Palaversa}, {Panahi}, {Pawlak}, {Piersimoni}, {Pineau}, {Plachy},
  {Plum}, {Poggio}, {Poujoulet}, {Pr{\v{s}}a}, {Pulone}, {Racero}, {Ragaini},
  {Rambaux}, {Ramos-Lerate}, {Regibo}, {Reyl{\'e}}, {Riclet}, {Ripepi}, {Riva},
  {Rivard}, {Rixon}, {Roegiers}, {Roelens}, {Romero-G{\'o}mez}, {Rowell},
  {Royer}, {Ruiz-Dern}, {Sadowski}, {Sagrist{\`a} Sell{\'e}s}, {Sahlmann},
  {Salgado}, {Salguero}, {Sanna}, {Santana-Ros}, {Sarasso}, {Savietto},
  {Schultheis}, {Sciacca}, {Segol}, {Segovia}, {S{\'e}gransan}, {Shih},
  {Siltala}, {Silva}, {Smart}, {Smith}, {Solano}, {Solitro}, {Sordo}, {Soria
  Nieto}, {Souchay}, {Spagna}, {Spoto}, {Stampa}, {Steele},
  {Steidelm{\"u}ller}, {Stephenson}, {Stoev}, {Suess}, {Surdej}, {Szabados},
  {Szegedi-Elek}, {Tapiador}, {Taris}, {Tauran}, {Taylor}, {Teixeira},
  {Terrett}, {Teyssand ier}, {Thuillot}, {Titarenko}, {Torra Clotet}, {Turon},
  {Ulla}, {Utrilla}, {Uzzi}, {Vaillant}, {Valentini}, {Valette}, {van Elteren},
  {Van Hemelryck}, {van Leeuwen}, {Vaschetto}, {Vecchiato}, {Veljanoski},
  {Viala}, {Vicente}, {Vogt}, {von Essen}, {Voss}, {Votruba}, {Voutsinas},
  {Walmsley}, {Weiler}, {Wertz}, {Wevers}, {Wyrzykowski}, {Yoldas},
  {{\v{Z}}erjal}, {Ziaeepour}, {Zorec}, {Zschocke}, {Zucker}, {Zurbach}, \&
  {Zwitter}}]{GaiaCollaboration2018a}
{Gaia Collaboration}, {Brown}, A.~G.~A., {Vallenari}, A., {et~al.} 2018, \aap,
  616, A1

\bibitem[{{Gaia Collaboration} {et~al.}(2016){Gaia Collaboration}, {Prusti},
  {de Bruijne}, {Brown}, {Vallenari}, {Babusiaux}, {Bailer-Jones}, {Bastian},
  {Biermann}, {Evans}, \& et~al.}]{GaiaCollaboration2016a}
{Gaia Collaboration}, {Prusti}, T., {de Bruijne}, J.~H.~J., {et~al.} 2016,
  \aap, 595, A1

\bibitem[{{Gallart} {et~al.}(2019){Gallart}, {Bernard}, {Brook}, {Ruiz-Lara},
  {Cassisi}, {Hill}, \& {Monelli}}]{Gallart2019a}
{Gallart}, C., {Bernard}, E.~J., {Brook}, C.~B., {et~al.} 2019, Nature
  Astronomy, 3, 407

\bibitem[{Gratton {et~al.}(2003)Gratton, Carretta, Desidera, Lucatello, Mazzei,
  \& Barbieri}]{Gratton2003a}
Gratton, R.~G., Carretta, E., Desidera, S., {et~al.} 2003, \aap, 406, 131

\bibitem[{Green {et~al.}(2019)Green, Schlafly, Zucker, Speagle, \&
  Finkbeiner}]{Green2019a}
Green, G.~M., Schlafly, E., Zucker, C., Speagle, J.~S., \& Finkbeiner, D. 2019,
  \apj, 887, 93

\bibitem[{{Grillmair}(2006)}]{Grillmair2006a}
{Grillmair}, C.~J. 2006, \apjl, 645, L37

\bibitem[{{Gustafsson} {et~al.}(2008){Gustafsson}, {Edvardsson}, {Eriksson},
  {J{\o}rgensen}, {Nordlund}, \& {Plez}}]{Gustafsson2008}
{Gustafsson}, B., {Edvardsson}, B., {Eriksson}, K., {et~al.} 2008, \aap, 486,
  951

\bibitem[{Hasselquist {et~al.}(2021)Hasselquist, Hayes, Lian, Weinberg,
  Zasowski, Horta, Beaton, Feuillet, Garro, Gallart, Smith, Holtzman, Minniti,
  Shetrone, J{\"o}nsson, Cioni, Fillingham, Cunha, O{\'C}onnell,
  Fern{\'a}ndez-Trincado, Mu{\~n}oz, Schiavon, Almeida, Anguiano, Beers,
  Bizyaev, Brownstein, Cohen, Frinchaboy, Garc{\'\i}a-Hern{\'a}ndez, Geisler,
  Lane, Majewski, Nidever, Nitschelm, Povick, Price-Whelan, Roman-Lopes,
  Rosado, Sobeck, Stringfellow, Valenzuela, Villanova, \&
  Vincenzo}]{Hasselquist2021a}
Hasselquist, S., Hayes, C.~R., Lian, J., {et~al.} 2021, arXiv e-prints,
  arXiv:2109.05130

\bibitem[{{Haywood} {et~al.}(2018){Haywood}, {Di Matteo}, {Lehnert}, {Snaith},
  {Khoperskov}, \& {G{\'o}mez}}]{Haywood2018a}
{Haywood}, M., {Di Matteo}, P., {Lehnert}, M.~D., {et~al.} 2018, \apj, 863, 113

\bibitem[{{Helmi} {et~al.}(2018){Helmi}, {Babusiaux}, {Koppelman}, {Massari},
  {Veljanoski}, \& {Brown}}]{Helmi2018a}
{Helmi}, A., {Babusiaux}, C., {Koppelman}, H.~H., {et~al.} 2018, \nat, 563, 85

\bibitem[{Hendricks {et~al.}(2014)Hendricks, Koch, Lanfranchi, Boeche, Walker,
  Johnson, Pe{\~n}arrubia, \& Gilmore}]{Hendricks2014a}
Hendricks, B., Koch, A., Lanfranchi, G.~A., {et~al.} 2014, \apj, 785, 102

\bibitem[{{Hill} {et~al.}(2019){Hill}, {Sk{\'u}lad{\'o}ttir}, {Tolstoy},
  {Venn}, {Shetrone}, {Jablonka}, {Primas}, {Battaglia}, {de Boer},
  {Fran{\c{c}}ois}, {Helmi}, {Kaufer}, {Letarte}, {Starkenburg}, \&
  {Spite}}]{Hill2019a}
{Hill}, V., {Sk{\'u}lad{\'o}ttir}, {\'A}., {Tolstoy}, E., {et~al.} 2019, \aap,
  626, A15

\bibitem[{{Ibata} {et~al.}(1994){Ibata}, {Gilmore}, \& {Irwin}}]{Ibata1994a}
{Ibata}, R.~A., {Gilmore}, G., \& {Irwin}, M.~J. 1994, \nat, 370, 194

\bibitem[{{Ishigaki} {et~al.}(2012){Ishigaki}, {Chiba}, \&
  {Aoki}}]{Ishigaki2012}
{Ishigaki}, M.~N., {Chiba}, M., \& {Aoki}, W. 2012, \apj, 753, 64

\bibitem[{Ji {et~al.}(2020)Ji, Li, Hansen, Casey, Koposov, Pace, Mackey, Lewis,
  Simpson, Bland-Hawthorn, Cullinane, Da~Costa, Hattori, Martell, Kuehn, Erkal,
  Shipp, Wan, \& Zucker}]{Ji2020a}
Ji, A.~P., Li, T.~S., Hansen, T.~T., {et~al.} 2020, \aj, 160, 181

\bibitem[{King(1997)}]{King1997a}
King, J.~R. 1997, \aj, 113, 2302

\bibitem[{Kirby {et~al.}(2011)Kirby, Cohen, Smith, Majewski, Sohn, \&
  Guhathakurta}]{Kirby2011a}
Kirby, E.~N., Cohen, J.~G., Smith, G.~H., {et~al.} 2011, \apj, 727, 79

\bibitem[{Kirby {et~al.}(2019)Kirby, Xie, Guo, de~los Reyes, Bergemann,
  Kovalev, Shen, Piro, \& McWilliam}]{Kirby2019a}
Kirby, E.~N., Xie, J.~L., Guo, R., {et~al.} 2019, \apj, 881, 45

\bibitem[{{Koppelman} {et~al.}(2018){Koppelman}, {Helmi}, \&
  {Veljanoski}}]{Koppelman2018a}
{Koppelman}, H., {Helmi}, A., \& {Veljanoski}, J. 2018, \apjl, 860, L11

\bibitem[{{Koppelman} {et~al.}(2019){Koppelman}, Helmi, Massari, Price-Whelan,
  \& Starkenburg}]{Koppelman2019a}
{Koppelman}, H.~H., Helmi, A., Massari, D., Price-Whelan, A.~M., \&
  Starkenburg, T.~K. 2019, \aap, 631, L9

\bibitem[{Kroupa \& Weidner(2003)}]{Kroupa2003a}
Kroupa, P. \& Weidner, C. 2003, \apj, 598, 1076

\bibitem[{Lallement {et~al.}(2019)Lallement, Babusiaux, Vergely, Katz, Arenou,
  Valette, Hottier, \& Capitanio}]{Lallement2019a}
Lallement, R., Babusiaux, C., Vergely, J.~L., {et~al.} 2019, \aap, 625, A135

\bibitem[{{Lemasle} {et~al.}(2014){Lemasle}, {de Boer}, {Hill}, {Tolstoy},
  {Irwin}, {Jablonka}, {Venn}, {Battaglia}, {Starkenburg}, {Shetrone},
  {Letarte}, {Fran{\c c}ois}, {Helmi}, {Primas}, {Kaufer}, \&
  {Szeifert}}]{Lemasle2014}
{Lemasle}, B., {de Boer}, T.~J.~L., {Hill}, V., {et~al.} 2014, \aap, 572, A88

\bibitem[{{Lemasle} {et~al.}(2012){Lemasle}, {Hill}, {Tolstoy}, {Venn},
  {Shetrone}, {Irwin}, {de Boer}, {Starkenburg}, \& {Salvadori}}]{Lemasle2012}
{Lemasle}, B., {Hill}, V., {Tolstoy}, E., {et~al.} 2012, \aap, 538, A100

\bibitem[{Lim {et~al.}(2021)Lim, Koch-Hansen, Hansen, L{\'e}pine, Marshall,
  Wilkinson, \& Pe{\~n}arrubia}]{Lim2021a}
Lim, D., Koch-Hansen, A.~J., Hansen, C.~J., {et~al.} 2021, \aap, 655, A26

\bibitem[{{Lind} {et~al.}(2011){Lind}, {Asplund}, {Barklem}, \&
  {Belyaev}}]{Lind2011a}
{Lind}, K., {Asplund}, M., {Barklem}, P.~S., \& {Belyaev}, A.~K. 2011, \aap,
  528, A103

\bibitem[{L{\"o}vdal {et~al.}(2022)L{\"o}vdal, Ruiz-Lara, Koppelman, Matsuno,
  Dodd, \& Helmi}]{Loevdal2022a}
L{\"o}vdal, S.~S., Ruiz-Lara, T., Koppelman, H.~H., {et~al.} 2022, arXiv
  e-prints, arXiv:2201.02404

\bibitem[{{Mackereth} {et~al.}(2019){Mackereth}, {Schiavon}, {Pfeffer},
  {Hayes}, {Bovy}, {Anguiano}, {Allende Prieto}, {Hasselquist}, {Holtzman},
  {Johnson}, {Majewski}, {O'Connell}, {Shetrone}, {Tissera}, \&
  {Fern{\'a}ndez-Trincado}}]{Mackereth2019a}
{Mackereth}, J.~T., {Schiavon}, R.~P., {Pfeffer}, J., {et~al.} 2019, \mnras,
  482, 3426

\bibitem[{Malhan {et~al.}(2018)Malhan, Ibata, \& Martin}]{Malhan2018a}
Malhan, K., Ibata, R.~A., \& Martin, N.~F. 2018, \mnras, 481, 3442

\bibitem[{Matsuno {et~al.}(2021{\natexlab{a}})Matsuno, Aoki, Casagrande,
  Ishigaki, Shi, Takata, Xiang, Yong, Li, Suda, Xing, \& Zhao}]{Matsuno2021a}
Matsuno, T., Aoki, W., Casagrande, L., {et~al.} 2021{\natexlab{a}}, \apj, 912,
  72

\bibitem[{{Matsuno} {et~al.}(2019){Matsuno}, {Aoki}, \& {Suda}}]{Matsuno2019a}
{Matsuno}, T., {Aoki}, W., \& {Suda}, T. 2019, \apjl, 874, L35

\bibitem[{Matsuno {et~al.}(2021{\natexlab{b}})Matsuno, Hirai, Tarumi,
  Hotokezaka, Tanaka, \& Helmi}]{Matsuno2021b}
Matsuno, T., Hirai, Y., Tarumi, Y., {et~al.} 2021{\natexlab{b}}, \aap, 650,
  A110

\bibitem[{{McMillan}(2017)}]{McMillan2017a}
{McMillan}, P.~J. 2017, \mnras, 465, 76

\bibitem[{Monty {et~al.}(2020)Monty, Venn, Lane, Lokhorst, \&
  Yong}]{Monty2020a}
Monty, S., Venn, K.~A., Lane, J. M.~M., Lokhorst, D., \& Yong, D. 2020, \mnras,
  497, 1236

\bibitem[{{Myeong} {et~al.}(2018){Myeong}, {Evans}, {Belokurov}, {Sanders}, \&
  {Koposov}}]{Myeong2018c}
{Myeong}, G.~C., {Evans}, N.~W., {Belokurov}, V., {Sanders}, J.~L., \&
  {Koposov}, S.~E. 2018, \apjl, 856, L26

\bibitem[{{Myeong} {et~al.}(2019){Myeong}, {Vasiliev}, {Iorio}, {Evans}, \&
  {Belokurov}}]{Myeong2019a}
{Myeong}, G.~C., {Vasiliev}, E., {Iorio}, G., {Evans}, N.~W., \& {Belokurov},
  V. 2019, \mnras, 488, 1235

\bibitem[{Naidu {et~al.}(2020)Naidu, Conroy, Bonaca, Johnson, Ting, Caldwell,
  Zaritsky, \& Cargile}]{Naidu2020a}
Naidu, R.~P., Conroy, C., Bonaca, A., {et~al.} 2020, \apj, 901, 48

\bibitem[{Nidever {et~al.}(2020)Nidever, Hasselquist, Hayes, Hawkins, Povick,
  Majewski, Smith, Anguiano, Stringfellow, Sobeck, Cunha, Beers, Bestenlehner,
  Cohen, Garcia-Hernandez, J{\"o}nsson, Nitschelm, Shetrone, Lacerna,
  Allende~Prieto, Beaton, Dell'Agli, Fern{\'a}ndez-Trincado, Feuillet, Gallart,
  Hearty, Holtzman, Manchado, Mu{\~n}oz, O'Connell, \& Rosado}]{Nidever2020a}
Nidever, D.~L., Hasselquist, S., Hayes, C.~R., {et~al.} 2020, \apj, 895, 88

\bibitem[{Nissen \& Schuster(1997)}]{Nissen1997a}
Nissen, P.~E. \& Schuster, W.~J. 1997, \aap, 326, 751

\bibitem[{{Nissen} \& {Schuster}(2010)}]{Nissen2010}
{Nissen}, P.~E. \& {Schuster}, W.~J. 2010, \aap, 511, L10

\bibitem[{{Nissen} \& {Schuster}(2011)}]{Nissen2011}
{Nissen}, P.~E. \& {Schuster}, W.~J. 2011, \aap, 530, A15

\bibitem[{{Noguchi} {et~al.}(2002){Noguchi}, {Aoki}, {Kawanomoto}, {Ando},
  {Honda}, {Izumiura}, {Kambe}, {Okita}, {Sadakane}, {Sato}, {Tajitsu},
  {Takada-Hidai}, {Tanaka}, {Watanabe}, \& {Yoshida}}]{Noguchi2002}
{Noguchi}, K., {Aoki}, W., {Kawanomoto}, S., {et~al.} 2002, \pasj, 54, 855

\bibitem[{O'Malley {et~al.}(2017)O'Malley, McWilliam, Chaboyer, \&
  Thompson}]{Omalley2017a}
O'Malley, E.~M., McWilliam, A., Chaboyer, B., \& Thompson, I. 2017, \apj, 838,
  90

\bibitem[{Palla(2021)}]{Palla2021a}
Palla, M. 2021, \mnras, 503, 3216

\bibitem[{{Prantzos} {et~al.}(1990){Prantzos}, {Hashimoto}, \&
  {Nomoto}}]{Prantzos1990a}
{Prantzos}, N., {Hashimoto}, M., \& {Nomoto}, K. 1990, \aap, 234, 211

\bibitem[{{Ram{\'{\i}}rez} {et~al.}(2014){Ram{\'{\i}}rez}, {Mel{\'e}ndez},
  {Bean}, {Asplund}, {Bedell}, {Monroe}, {Casagrande}, {Schirbel}, {Dreizler},
  {Teske}, {Tucci Maia}, {Alves-Brito}, \& {Baumann}}]{Ramirez2014}
{Ram{\'{\i}}rez}, I., {Mel{\'e}ndez}, J., {Bean}, J., {et~al.} 2014, \aap, 572,
  A48

\bibitem[{Ramos {et~al.}(2020)Ramos, Mateu, Antoja, Helmi, Castro-Ginard,
  Balbinot, \& Carrasco}]{Ramos2020a}
Ramos, P., Mateu, C., Antoja, T., {et~al.} 2020, \aap, 638, A104

\bibitem[{Reggiani \& Mel{\'e}ndez(2018)}]{Reggiani2018a}
Reggiani, H. \& Mel{\'e}ndez, J. 2018, \mnras, 475, 3502

\bibitem[{{Reggiani} {et~al.}(2017){Reggiani}, {Mel{\'e}ndez}, {Kobayashi},
  {Karakas}, \& {Placco}}]{Reggiani2017a}
{Reggiani}, H., {Mel{\'e}ndez}, J., {Kobayashi}, C., {Karakas}, A., \&
  {Placco}, V. 2017, \aap, 608, A46

\bibitem[{{Reid} \& {Brunthaler}(2004)}]{Reid2004a}
{Reid}, M.~J. \& {Brunthaler}, A. 2004, \apj, 616, 872

\bibitem[{Ruiz-Dern {et~al.}(2018)Ruiz-Dern, Babusiaux, Arenou, Turon, \&
  Lallement}]{RuizDern2018a}
Ruiz-Dern, L., Babusiaux, C., Arenou, F., Turon, C., \& Lallement, R. 2018,
  \aap, 609, A116

\bibitem[{Ruiz-Lara {et~al.}(2022)Ruiz-Lara, Matsuno, Sofie~L{\"o}vdal, Helmi,
  Dodd, \& Koppelman}]{RuizLara2022a}
Ruiz-Lara, T., Matsuno, T., Sofie~L{\"o}vdal, S., {et~al.} 2022, arXiv
  e-prints, arXiv:2201.02405

\bibitem[{Sanders {et~al.}(2021)Sanders, Belokurov, \& Man}]{Sanders2021a}
Sanders, J.~L., Belokurov, V., \& Man, K. T.~F. 2021, \mnras, 506, 4321

\bibitem[{{Sch{\"o}nrich} {et~al.}(2010){Sch{\"o}nrich}, {Binney}, \&
  {Dehnen}}]{Schoenrich2010a}
{Sch{\"o}nrich}, R., {Binney}, J., \& {Dehnen}, W. 2010, \mnras, 403, 1829

\bibitem[{{Sneden}(1973)}]{Sneden1973}
{Sneden}, C. 1973, \apj, 184, 839

\bibitem[{{Stephens} \& {Boesgaard}(2002)}]{Stephens2002}
{Stephens}, A. \& {Boesgaard}, A.~M. 2002, \aj, 123, 1647

\bibitem[{{Suda} {et~al.}(2017){Suda}, {Hidaka}, {Aoki}, {Katsuta}, {Yamada},
  {Fujimoto}, {Ohtani}, {Masuyama}, {Noda}, \& {Wada}}]{Suda2017a}
{Suda}, T., {Hidaka}, J., {Aoki}, W., {et~al.} 2017, \pasj, 69, 76

\bibitem[{{Suda} {et~al.}(2008){Suda}, {Katsuta}, {Yamada}, {Suwa}, {Ishizuka},
  {Komiya}, {Sorai}, {Aikawa}, \& {Fujimoto}}]{Suda2008}
{Suda}, T., {Katsuta}, Y., {Yamada}, S., {et~al.} 2008, \pasj, 60, 1159

\bibitem[{{Suda} {et~al.}(2011){Suda}, {Yamada}, {Katsuta}, {Komiya},
  {Ishizuka}, {Aoki}, \& {Fujimoto}}]{Suda2011}
{Suda}, T., {Yamada}, S., {Katsuta}, Y., {et~al.} 2011, \mnras, 412, 843

\bibitem[{{Tajitsu} {et~al.}(2012){Tajitsu}, {Aoki}, \&
  {Yamamuro}}]{Tajitsu2012a}
{Tajitsu}, A., {Aoki}, W., \& {Yamamuro}, T. 2012, \pasj, 64, 77

\bibitem[{Theler {et~al.}(2020)Theler, Jablonka, Lucchesi, Lardo, North, Irwin,
  Battaglia, Hill, Tolstoy, Venn, Helmi, Kaufer, Primas, \&
  Shetrone}]{Theler2020a}
Theler, R., Jablonka, P., Lucchesi, R., {et~al.} 2020, \aap, 642, A176

\bibitem[{{Tolstoy} {et~al.}(2009){Tolstoy}, {Hill}, \& {Tosi}}]{Tolstoy2009}
{Tolstoy}, E., {Hill}, V., \& {Tosi}, M. 2009, \araa, 47, 371

\bibitem[{{Vasiliev}(2019)}]{Vasiliev2019a}
{Vasiliev}, E. 2019, \mnras, 482, 1525

\bibitem[{{Venn} {et~al.}(2004){Venn}, {Irwin}, {Shetrone}, {Tout}, {Hill}, \&
  {Tolstoy}}]{Venn2004a}
{Venn}, K.~A., {Irwin}, M., {Shetrone}, M.~D., {et~al.} 2004, \aj, 128, 1177

\bibitem[{{Vogt} {et~al.}(1994){Vogt}, {Allen}, {Bigelow}, {Bresee}, {Brown},
  {Cantrall}, {Conrad}, {Couture}, {Delaney}, {Epps}, {Hilyard}, {Hilyard},
  {Horn}, {Jern}, {Kanto}, {Keane}, {Kibrick}, {Lewis}, {Osborne},
  {Pardeilhan}, {Pfister}, {Ricketts}, {Robinson}, {Stover}, {Tucker}, {Ward},
  \& {Wei}}]{Vogt1994}
{Vogt}, S.~S., {Allen}, S.~L., {Bigelow}, B.~C., {et~al.} 1994, in \procspie,
  Vol. 2198, Instrumentation in Astronomy VIII, ed. D.~L. {Crawford} \& E.~R.
  {Craine}, 362

\bibitem[{{Yamada} {et~al.}(2013){Yamada}, {Suda}, {Komiya}, {Aoki}, \&
  {Fujimoto}}]{Yamada2013}
{Yamada}, S., {Suda}, T., {Komiya}, Y., {Aoki}, W., \& {Fujimoto}, M.~Y. 2013,
  \mnras, 436, 1362

\bibitem[{Yuan {et~al.}(2020)Yuan, Myeong, Beers, Evans, Lee, Banerjee, Gudin,
  Hattori, Li, Matsuno, Placco, Smith, Whitten, \& Zhao}]{Yuan2020a}
Yuan, Z., Myeong, G.~C., Beers, T.~C., {et~al.} 2020, \apj, 891, 39

\bibitem[{{Zolotov} {et~al.}(2010){Zolotov}, {Willman}, {Brooks}, {Governato},
  {Hogg}, {Shen}, \& {Wadsley}}]{Zolotov2010}
{Zolotov}, A., {Willman}, B., {Brooks}, A.~M., {et~al.} 2010, \apj, 721, 738

\end{thebibliography}
\end{document}